\documentclass{lmcs}
\pdfoutput=1
\usepackage[utf8]{inputenc}
\usepackage[T1]{fontenc}
\usepackage[hidelinks]{hyperref}
\usepackage{amsmath,amssymb}
\usepackage{cleveref}
\usepackage{tikz-cd}
\usepackage{macros}
\usepackage{amsmath}
\usepackage{mathtools}
\usepackage{verbatim}
\usepackage{utf8}
\usepackage{theorems}

\title{Delooping presented groups in homotopy type theory}
\author{Camil Champin}[a]
\author{Samuel Mimram\lmcsorcid{0000-0002-0767-2569}}[b]
\author{Émile Oleon\lmcsorcid{0009-0001-8398-2577}}[b]

\address{École Normale Superieure de Lyon}
\email{camil.champin@ens-lyon.fr}
\address{LIX, CNRS, École polytechnique, Institut Polytechnique de Paris, 91120 Palaiseau, France}
\email{samuel.mimram@polytechnique.edu, emile.oleon@polytechnique.edu}

\crefformat{equation}{(#2#1#3)}
\Crefname{figure}{Figure}{Figures}
\begin{document}
\maketitle

\begin{abstract}
  Homotopy type theory is a logical setting based on Martin-Löf type theory in which geometric constructions and proofs can be carried out synthetically. Here, types can be interpreted as spaces up to homotopy, and proofs as homotopy-invariant constructions. In this context, the loop spaces of pointed connected groupoids provide a natural representation of groups, and every group can be realized as the loop space of such a type, which is then called a \emph{delooping} of the group. There are two main methods for constructing a delooping of an arbitrary group~$G$. The first describes it as a pointed higher inductive type, while the second takes the connected component of the principal $G$-torsor in the type of sets equipped with a $G$-action. We show that, when a presentation, or even just a generating set, is known for the group, simpler variants of these constructions can be used to build deloopings. The resulting types are more amenable to computation and lead to simpler metatheoretic reasoning. Finally, we develop a type-theoretic notion of 2-polygraph for manipulating higher inductive types such as those arising in the description of deloopings. This allows us to investigate a construction of the Cayley graph of a generated group and to show that it encodes the relations of the group, as well as a Cayley complex encoding relations between relations. Many of the developments in this article have been formalized in the cubical version of the Agda proof assistant.

\end{abstract}


\section*{Introduction}
Homotopy type theory was introduced around 2010~\cite{hottbook}, based on Martin-Löf type theory~\cite{martin1984intuitionistic}. It starts from the idea that types in logic should be interpreted not only as sets, as traditionally done in the semantics of logic, but rather as \emph{spaces} considered up to homotopy. Namely, the identities between two elements of a type can be thought of as paths between points corresponding to the elements, identities on identities as homotopies between paths, and so on. Moreover, this correspondence can be made to work precisely, by postulating the \emph{univalence axiom}~\cite{kapulkin2021simplicial}, which states that identities between types coincide with equivalences. This opens the way to the implementation of geometric constructions in a synthetic manner, by performing operations on types, which will semantically correspond to the desired geometrical operations.
In this setting, we are interested in providing ways to construct concise models of groups that allow simple proofs and simplify metatheoretic reasoning.

\paragraph{Delooping groups}
Following a well-known construction due to Poincaré at the end of the 19th
century~\cite{poincare1895analysis}, to any type~$A$ which is pointed, \ie
equipped with a distinguished element~$\pt$, we can associate its
\emph{fundamental group} $\pi_1A$, defined as $\strunc{\pt=\pt}$, whose elements are the
homotopy classes of paths from $\pt$ to itself, with composition given by
concatenation and identity by the constant path. Moreover, when the type~$A$ is
a groupoid, in the sense that any two homotopies between paths are homotopic,
this fundamental group coincides with the \emph{loop space}
$\Loop A$, which is $\pt=\pt$, \ie defined similarly but without quotienting paths up
to homotopy.
Once this observation has been made, it is natural to wonder whether every group~$G$
arises as the loop space of some groupoid. It turns out that this is the case:
to every group one can associate a pointed connected groupoid type $\B G$,
called its \emph{delooping}, whose loop space is~$G$. Moreover, there is
essentially only one such type, thus justifying the notation.


\paragraph{Internal and external points of view}
The delooping construction, which can be found in various places~\cite{symmetry,
  buchholtz2023central}, and will be recalled in the article, induces an
equivalence between the type of groups and the type of pointed connected
groupoids (\cref{internal-external-equivalence}).
This thus provides us with two alternative descriptions of groups in homotopy
type theory. The one as (loop spaces of) pointed connected groupoids can be
thought of as an \emph{internal} one, since the structure is deduced from the
types without imposing further axioms; by contrast, the more traditional one as
sets equipped with multiplication and unit operations
is rather an \emph{external} one (some also use the terminology \emph{concrete} and \emph{abstract} instead of internal and external~\cite{symmetry}, and this also respectively corresponds to having a \emph{shallow embedding} or a \emph{deep embedding} of the notion of group).
We should also say here that pointed connected types (which are not necessarily
groupoids) can be thought of as higher versions of groups, where the axioms only
hold up to higher identities which are themselves coherent, and so on.


\paragraph{Two ways to construct deloopings}
Two generic ways are currently known to construct the delooping $\B G$
of a group~$G$, which we both refine in this article.
The first one is the \emph{torsor} construction which originates in algebraic
topology~\cite{demazure1970groupes} and can be adapted in homotopy type
theory~\cite{symmetry,buchholtz2023central,warn2023eilenberg}. One can consider
the type of \emph{$G$-sets}, which are sets equipped with an action of~$G$. Among
those, there is a canonical one, called the \emph{principal
  $G$-torsor}~$\ptorsor G$, which arises from the action of the group~$G$ on
itself by left multiplication.  It can be shown that the loop space of the type
of $G$-sets, pointed on the principal $G$-torsor~$\ptorsor G$, is the group~$G$.
Moreover, if one restricts the type of $G$-sets to the connected component of
the principal $G$-torsor, one obtains the type of \emph{$G$-torsors}, which is a
delooping of~$G$.

The second one is a particular case of the definition of \emph{Eilenberg-MacLane spaces} in homotopy type theory due to Finster and Licata~\cite{licata2014eilenberg}. It consists in constructing $\B G$ as a higher inductive type with one point, one loop for each element of~$G$, one identity for each entry in the multiplication table of~$G$, and then truncating the resulting type as a groupoid. One can imagine that the resulting space has the right loop space ``by construction'', although the formal proof is non-trivial.

\paragraph{Torsors for generated groups}
In this article, we refine the above two constructions in order to obtain versions that are ``simpler'' (in the sense that they have fewer constructors or require introducing less material) when a presentation by generators and relations is known for the group.
For the construction based on $G$-torsors, we show that a simpler definition can be achieved when a generating set~$X$ is known for~$G$. Namely, we show that one can perform essentially the same construction, but replacing $G$-sets by what we call here $X$-sets (\cref{generated-delooping}), where we only need to consider the action of the generators (as opposed to the whole group).
As an illuminating example, consider the case $G=\Z$, whose delooping is known to be the circle $\B\Z=\S1$. The type~$\Endomorphisms\eqdef\Sigma(X:\U).(X\to X)$ of all endomorphisms, on any type, contains, as a particular element, the successor function~$s:\Z\to\Z$. Our results imply that the connected component of~$s$ in~$\Endomorphisms$ is a delooping of~$\Z$. This description is arguably simpler than that of $\Z$-torsors: indeed, morphisms of $\Z$-sets are required to preserve the action of every element of~$\Z$, while morphisms in~$\Endomorphisms$ are only required to preserve the action of~$1$ (which corresponds to the successor).
The above description is the one used in UniMath to define the circle $\S1$~\cite{bezem2021construction}: the reason they use it instead of the traditional one~\cite{hottbook} is that they do not allow themselves to use higher inductive types, because these are not yet entirely clear from a metatheoretic point of view (there is no general definition, even though there are proposals~\cite{lumsdaine2020semantics}, and the semantics of type theory~\cite{kapulkin2021simplicial} has not been fully worked out in their presence).
Our results thus provide an abstract explanation of why this construction works and a generic way to define many more deloopings without resorting to higher inductive types, if one prefers not to use them.

\paragraph{Higher inductive types for presented groups}
For the second construction of the delooping, as a higher inductive type, we show that~$\B G$ can be constructed as the higher inductive type generated by one point, one loop for each generator of the presentation (as opposed to every element of the group), one identity for each relation of the presentation, and a groupoid truncation (\cref{presented-delooping}). This has the advantage of resulting in types that are simpler to define, require handling fewer cases when reasoning by induction, and match the usual combinatorial descriptions of groups. It thus allows one to carry out, in a synthetic way, proofs close to traditional ones in group theory, such as those based on Tietze transformations.
Moreover, we claim that the traditional methods based on rewriting~\cite{polygraphs,groves2006rewriting} in order to compute invariants such as homology or coherence can be applied to those. Namely, a first important step in this direction was obtained by Kraus and von Raumer's adaptation of Squier's coherence theorem in homotopy type theory~\cite{kraus2022rewriting}.


\paragraph{Polygraphs}
In order to describe precisely the higher inductive types involved in the previous construction, and to reason internally about them, we recall the definition of 2-polygraphs, which originates in the study of rewriting and $\omega$-categories~\cite{polygraphs} and was recently adapted to the setting of homotopy type theory in~\cite{kraus2022rewriting}. We further develop their theory by introducing the notions of types generated and presented by a 2-polygraph, and show that the delooping $\B G$ constructed with higher inductive types precisely corresponds to a presented type. Finally, in order to demonstrate the applicability of traditional group-theoretic techniques, we develop a notion of Tietze transformation for 2-polygraphs and show that such transformations preserve the presented type (\cref{tietze-invariant}).

\paragraph{Cayley graphs}
As another aspect of our study of generated groups in homotopy type theory, we provide a pleasant abstract description of Cayley graphs, a well-known construction in group theory~\cite{cayley1878desiderata,lyndon1977combinatorial}. We show that, given a group~$G$ with a set~$X$ of generators, the Cayley graph can be obtained as the kernel of the canonical map $\B\freegroup{X}\to\B G$, where~$\freegroup X$ is the free group on~$X$ (\cref{flattening-cayley}), and admits a natural description by a 2-polygraph. This establishes these graphs as a measure of the difference between deloopings and deloopings of free groups, and suggests higher-dimensional versions of them: we also introduce Cayley complexes, which measure relations between relations (\cref{cayley-complex-fiber-sequence}).

\paragraph{Formalization}
Many of the results presented in this article have been formalized in the cubical variant of the Agda proof assistant~\cite{vezzosi2021cubical} using the ``standard library'' which has been developed for it~\cite{cubicalagda}. Our developments are publicly available~\cite{git}, and we provide pointers to the formalized results when available.


\paragraph{Plan of the paper}
We begin by briefly recalling the fundamental notions of homotopy type theory that will be used throughout the paper (\cref{hott}), as well as the notion of delooping for a group (\cref{delooping}).
We first present the construction of deloopings based on the torsor construction (\cref{torsor-delooping}) and show how it can be simplified when a generating set is known for the group (\cref{generated-torsors}).
We then present the other approach for defining deloopings of groups using higher inductive types, and explain how those can be simplified when a presentation is known for the group (\cref{delooping-hit}).
We introduce the structure of 2-polygraph and show that it allows encoding this notion of delooping of group, while supporting internal manipulations (\cref{polygraphs}).
Finally, we investigate the construction of Cayley graphs and complexes in homotopy type theory (\cref{cayley}).
We conclude by presenting possible extensions of this work (\cref{conclusion}).

\paragraph{Acknowledgments}
We would like to thank Dan Christensen as well as an anonymous reviewer for useful comments on early drafts of this article, which is an extended version of~\cite{generated}.



\section{Homotopy type theory}
\label{hott}
In this section, we recall the main notions from homotopy type theory that will be used in the rest of the paper. A detailed presentation of the subject can be found in the reference books~\cite{hottbook,rijke2022introduction}.

\paragraph{Universe}
We write $\U$ for the \emph{universe}, \ie the large type of all small types, which we assume to be closed under dependent sums and products.
We write $0$, $1$, $2$ (and so on) for the initial type, the terminal type and the type of booleans.
Given a type~$A$ and a type family $B:A\to\U$, we write $\Pi(x:A).B\,x$ or $(x:A)\to B\,x$ or $\Pi A.B$ for $\Pi$-types, and $A\to B$ for the case where~$B$ does not depend on~$A$. Similarly, we write $\Sigma(x:A).B\,x$ or $\Sigma A.B$ for $\Sigma$-types, and $A\times B$ for the non-dependent version. The two projections from a $\Sigma$-type are respectively written $\fst:\Sigma A.B\to A$ and $\snd:(x:\Sigma A.B)\to B(\fst x)$.
Given type families $B:A\to\U$ and $B':A'\to\U$, a pair of maps $f:A\to A'$ and $g:(x:A)\to B\,x\to B'(f\,x)$ canonically induces a map denoted $\Sigma f.g:\Sigma A.B\to\Sigma A'.B'$.

\paragraph{Paths}
Given a type $A:\U$ and two elements $a,b:A$, we write $a=_A b$ (or simply $a=b$) for the type of \emph{identities}, or \emph{paths}, between~$a$ and~$b$: its elements are proofs of equality between $a$ and $b$. In particular, for any element $a:A$, the type $a=a$ contains the term $\refl[a]$ witnessing reflexivity of equality.
We sometimes write $t\eqdef u$ to indicate that $t$ and $u$ are equal by definition.
The elimination principle of identities, \aka \emph{path induction} and often denoted~$J$, states that, given $a:A$, in order to show a property $P:(x:A)\to(a=x)\to\U$ for every $x:A$ and $p:a=x$, it is enough to show it in the case where $x\defd a$ and $p\defd\refl[a]$.
%
%
Given paths $p:x=y$ and $q:y=z$, we write $p\pcomp q$ for their concatenation and $\sym p:y=x$ for the inverse of~$p$.

Given a type~$A$ and a type family $B:A\to\U$, a path $p:x=y$ in~$A$ induces a function $\subst Bp:B(x)\to B(y)$ expressing the fact that equality is \emph{substitutive}. As a special case, any path $p:A=B$ between two types~$A,B:\U$ induces a function $\transport p:A\to B$, called the \emph{transport} along~$p$, as well as an inverse function $\transportinv p:B\to A$. Finally, given a function $f:A\to B$, any path $p:x=y$ in~$A$ induces a path $\Cong f(p): f(x) = f(y)$ expressing the fact that equality is a \emph{congruence}.

\paragraph{Pointed types}
A \emph{pointed} type consists of a type~$A$ together with a distinguished element, often written~$\pt[A]$, or even $\pt$, and sometimes left implicit. We write $\pType\defd\Sigma(A:\Type).A$ for the type of pointed types. A \emph{pointed map} $f:A\to B$ between pointed types is a map between the underlying types equipped with an identification $\pt[f]:f\pt[A]=\pt[B]$ and we write $A\pto B$ for the type of pointed maps.

\paragraph{Higher inductive types}
Many functional programming languages allow the definition of inductive types, which are freely generated by constructors. For instance, the type~$\S0$ of booleans is generated by two elements (true and false).
In the context of homotopy type theory, languages such as cubical Agda~\cite{vezzosi2021cubical} feature a useful generalization of such types, called \emph{higher inductive types}. In addition to traditional constructors for elements of the type, they allow constructors for equalities between elements. For instance, the type corresponding to the circle $\S1$ can be defined as the type generated by two points $a$ and $b$ and two equalities $p,q:a=b$ between those points:
\begin{center}
\begin{minipage}{0.4\linewidth}
\begin{verbatim}
data S¹ : Type where
  a b : S¹
  p q : a ≡ b
\end{verbatim}
\end{minipage}
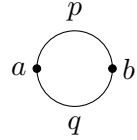
\begin{minipage}{0.4\linewidth}
  \[
    \begin{tikzpicture}[baseline=(current bounding box.center)]
      \filldraw (-.5,0) circle (0.05) node[left] {$a$};
      \filldraw (.5,0) circle (0.05) node[right] {$b$};
      \draw (0,0) circle (.5);
      \draw (0,.5) node[above]{$p$};
      \draw (0,-.5) node[below]{$q$};
    \end{tikzpicture}
  \]
\end{minipage}
\end{center}
(on the left is the definition of the higher inductive type in Agda, and on the right its geometric interpretation is depicted). Higher-dimensional spheres~$\S n$, for $n$ a natural number, can be defined in a similar way.

\paragraph{Univalence}
\label{equivalence}
A map $f:A\to B$ is an \emph{equivalence} when it admits both a left and a right inverse, \ie there are maps $g,g':B\to A$ together with identities $g\circ f=\id_A$ and $f\circ g'=\id_B$. In particular, every isomorphism is an equivalence. We write $A\equivto B$ for the type of equivalences from~$A$ to~$B$. The identity is clearly an equivalence and we thus have, by path induction, a canonical map
\[
  (A=B)\to(A\equivto B)
\]
for all types~$A$ and~$B$: the \emph{univalence axiom} states that this map is itself an equivalence.
In particular, every equivalence $A\equivto B$ induces a path $A=B$.
It is known that univalence implies the \emph{function extensionality} principle~\cite[Section~2.9]{hottbook}: given functions $f,g:A\to B$, if $f(x)=g(x)$ for any $x:A$ then $f=g$, and the expected generalization to dependent function types is also valid.

\paragraph{Homotopy levels}
A type~$A$ is \emph{contractible} when the type
\[
  \isContr(A)
  \quad\defd\quad
  \Sigma(x:A).(y:A)\to(x=y)
\]
is inhabited: this expresses the fact that we have a ``contraction point'' $x:A$, and a continuous family of paths from $x$ to every other point~$y$ in $A$.
A type~$A$ is a \emph{proposition} (\resp a \emph{set}, \resp a \emph{groupoid}) when the type $(x=y)$ is contractible (\resp a proposition, \resp a set) for every $x,y:A$. Intuitively, a contractible type is a point (up to homotopy), a proposition is either a point or empty, a set is a collection of points, and a groupoid is a space with no non-trivial 2-dimensional (or higher) structure.
For instance, consider the following four types corresponding to spheres of respective dimensions $-1$, $0$, $1$, and $2$:
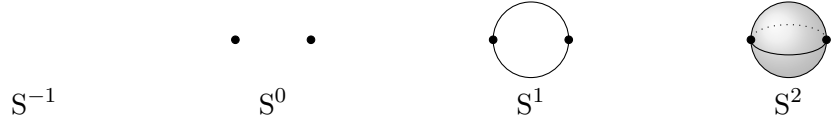
\[
  \begin{array}{c@{\qquad\qquad\qquad}c@{\qquad\qquad\qquad}c@{\qquad\qquad\qquad}c}
    &
    \begin{tikzpicture}
      \filldraw (0,0) circle (.05);
      \filldraw (1,0) circle (.05);
      \draw (.5,-.5);
    \end{tikzpicture}
    &
    \begin{tikzpicture}
      \filldraw (0,0) circle (.05);
      \filldraw (1,0) circle (.05);
      \draw (.5,0) circle (.5);
    \end{tikzpicture}
    &
    \begin{tikzpicture}
      \shade[ball color = gray!40, opacity = 0.4] (.5,0) circle (.5);
      \draw (.5,0) circle (.5);
      \draw (0,0) arc (180:360:.5 and 0.2);
      \draw[dotted] (1,0) arc (0:180:.5 and 0.2);
      \filldraw (0,0) circle (.05);
      \filldraw (1,0) circle (.05);
    \end{tikzpicture}
    \\
    \S{-1}&\S0&\S1&\S2
  \end{array}
  \]
The sphere $\S{-1}$ (\ie the empty type) is a proposition, the sphere $\S0$ (\ie the booleans) is a set and the sphere $\S1$ (\ie the circle)  is a groupoid, and the sphere $\S2$ is not a groupoid (because there is a 2-dimensional ``hole'' in it).
We write $\Set$ for the type of sets. Given a type~$A$, we write $\isSet(A)$ (\resp $\isGroupoid(A)$) for the predicate indicating that~$A$ is a set (\resp a groupoid), \ie we have
\[
  \isSet(A)
  \qdefd
  (x,y:A)\to(p,q:x=y)\to\isContr(p=q)
\]
and $\isGroupoid(A)$ is defined in a similar fashion.

\paragraph{Truncation}
Given a type $A$, its \emph{propositional truncation} turns it into a proposition in a universal way. It consists of a type~$\ptrunc{A}$, which is a proposition, equipped with a map~$\ptrunq{{-}}:A\to\ptrunc{A}$ such that, for any proposition~$B$, the map $(\ptrunc A\to B)\to(A\to B)$ induced by precomposition by~$\ptrunq{{-}}$ is an equivalence:
\[
  \begin{tikzcd}
    A\ar[d,"\ptrunq{{-}}"']\ar[r]&B\\
    \ptrunc A\ar[ur,dotted]
  \end{tikzcd}
\]
Intuitively, the type $\ptrunc A$ behaves like~$A$, except that we do not have access to its individual elements: the elimination principle for propositional truncation states that in order to construct an element of~$B$ from an element of~$\ptrunc A$, we can assume that we have an element of~$A$ only if~$B$ itself is a proposition. We say that a type is \emph{merely} inhabited when its propositional truncation is inhabited.
The \emph{set truncation} $\strunc A$ of a type~$A$ is defined similarly, as the universal way of turning~$A$ into a set, and we write $\strunq{x}$ for the image of~$x:A$ in the truncation; and we can similarly define the \emph{groupoid truncation}~$\gtrunc A$.
A type~$A$ is \emph{connected} when the type~$\strunc{A}$ is a proposition or, equivalently, when the path type $x=y$ is merely inhabited for every pair of elements~$x$ and~$y$.

\paragraph{Fibers}
Given a function $f:A\to B$ and an element $b:B$, the \emph{fiber} of~$f$ at $b$ is the type
\[
  \fib fb
  \qdefd
  \Sigma(a:A).(f\,a=b)
\]
A function $f$ is said to be \emph{surjective} when the type $(b:B)\to\ptrunc{\fib fb}$ is inhabited, \ie when every element of~$B$ merely has a preimage.
When $B$ is pointed, we write $\ker f\defd\fib f\pt$ for the fiber of $f$ at~$\pt$, called its \emph{kernel}. A sequence of composable arrows
\[
  \begin{tikzcd}
    F\ar[r]&A\ar[r,"f"]&B
  \end{tikzcd}
\]
is a \emph{fiber sequence} when $F$ is the kernel of~$f$ and the map $F\to A$ is the first projection, see~\cite[Section~8.4]{hottbook}.

\paragraph{Grothendieck duality}
\label{grothendieck-duality}
Any function $f:A\to B$ induces a type family $\fib f:B\to\U$ by taking the fibers of~$f$. Conversely, any type family~$F:B\to\U$ induces a function, namely the first projection $\fst:\Sigma B.F\to B$. In fact, these two constructions can be shown to define an equivalence, known as the \emph{Grothendieck duality}, between types over $B$ and type families indexed by~$B$~\cite[Section~4.8]{hottbook}:
\[
  \Sigma(A:\U).(A\to B)
  \quad\equivto\quad
  (B\to\U)
\]
Switching between the two points of view is often very useful.

\paragraph{The flattening lemma}
\label{flattening-coequalizers}
The \emph{flattening lemma} allows one to compute the total space of a type family indexed by a type that is a colimit. We recall it here in the case of coequalizers and pushouts, and refer to~\cite[Section~6.12]{hottbook} for a more detailed presentation and proof.
Suppose given a coequalizer
\[
  \begin{tikzcd}
    A\ar[r,shift left,"f"]\ar[r,shift right,"g"']&B\ar[r,dotted,"h"]&C
  \end{tikzcd}
\]
with $p:h\circ f=h\circ g$ witnessing the coequalization, and a type family $P:C\to\U$. Then the diagram
\[
  \begin{tikzcd}[sep=huge]
    \Sigma A.(P\circ h\circ f)\ar[r,shift left,"\Sigma f.(\lambda\_.\id{})"]\ar[r,shift right,"\Sigma g.e"']&\Sigma B.(P\circ h)\ar[r,dotted,"\Sigma h.(\lambda\_.\id{})"]&\Sigma C.P
  \end{tikzcd}
\]
is a coequalizer, where the map
\[
  e:(a:A)\to P\circ h\circ f(a)\to P\circ h\circ g(a)
\]
is induced by transport along~$p$, namely $e\,a\,x\defd\subst{P}{\happly pa}(x)$.
Note that there is a slight asymmetry: we could formulate a similar statement with $\Sigma A.(P\circ h\circ g)$ as the left object.

\label{flattening-pushouts}
The variant of the flattening lemma adapted to pushouts can be stated as follows. Consider a pushout square
\[
  \begin{tikzcd}
    X\ar[d,"f"']\ar[r,"g"]\ar[dr,phantom,pos=1,"\ulcorner"]&B\ar[d,"j"]\\
    A\ar[r,"i"']&A\sqcup_XB
  \end{tikzcd}
\]
with $p:i\circ f=j\circ g$ witnessing its commutativity, together with a type family $P:A\sqcup_XB\to\U$. Then the following square of total spaces is also a pushout
\[
  \begin{tikzcd}
    \Sigma X.(P\circ i\circ f)\ar[d,"\Sigma f.(\lambda\_.\id{})"']\ar[r,"\Sigma g.e"]&\Sigma B.(P\circ j)\ar[d,"\Sigma j.(\lambda\_.\id{})"]\\
    \Sigma A.(P\circ i)\ar[r,"\Sigma i.(\lambda\_.\id{})"']&\Sigma(A\sqcup_XB).P
  \end{tikzcd}
\]
where
\[
  e:(x:X)\to P\circ i\circ f(x)\to P\circ j\circ g(x)
\]
is the canonical map induced by~$p$, \ie $e(x)\defd\subst P{\happly px}$.


\section{Delooping groups}
\label{delooping}

\paragraph{The external point of view}
A \emph{group} consists of a set $A$, together with an operation $\gm:A\to A\to A$ (the \emph{multiplication}), an element $\ge:A$ (the \emph{unit}), and an operation $\gi:A\to A$ (the \emph{inverse}) such that the multiplication is associative, admits $\ge$ as a unit, and $\gi(x)$ is a two-sided inverse of any element $x:A$. We write $\Group$ for the type of all groups, and $G\toGroup H$ for the type of group morphisms between groups~$G$ and~$H$:
\begin{align*}
  \Group\qdefd
  \Sigma(A:\Set).&
  (\gm:A\to A\to A)\times(\ge:A)\times(\gi:A\to A)\times
  \\
  &((x,y,z:A)\to\gm(\gm(x,y),z)=\gm(x,\gm(y,z))\times
  \\
  &((x:A)\to(\gm(\ge,x)=x)\times(\gm(x,\ge)=x))\times
  \\
  &((x:A)\to(\gm(\gi(x),x)=\ge)\times(\gm(x,\gi(x))=\ge))
\end{align*}
In the following, we use the traditional notations for groups: given two elements $x,y:G$, we simply write $xy$ instead of $\gm(x,y)$, $1$ instead of $\ge$, and $x^{-1}$ instead of $\gi(x)$.

\paragraph{Loop types}
Given a pointed type $A$, its \emph{loop space}~$\Loop A$ is defined as the type of paths from $\pt$ to itself, which are called the \emph{loops} of~$A$:
\[
  \Loop A
  \qdefd
  (\pt=\pt)
\]
This operation extends to a map
\[
  \Loop:\pType\to\pType
\]
sending a pointed type~$A$ to the type~$\Loop A$ pointed by $\refl[\pt]$. It is moreover functorial in the sense that any pointed map $f:A\to B$ induces a map $\Loop f:\Loop A\to\Loop B$ sending $p:\Loop A$ to the loop $\Loop fp\eqdef\sym{\pt[f]}\pcomp\ap fp\pcomp\pt[f]$:
\[
  \begin{tikzcd}
    \pt[B]\ar[r,equals,"{\sym{\pt[f]}}"]&f\pt[A]\ar[r,equals,"\ap fp"]&f\pt[A]\ar[r,equals,"{\pt[f]}"]&\pt[B]
  \end{tikzcd}
\]
where the paths on the left and on the right are given by the fact that $f$ is pointed, see~\cite[Definition~8.4.2]{hottbook}. This construction is compatible with composition and identities, in the sense that $\Loop(g\circ f)=\Loop g\circ\Loop f$ and $\Loop\id=\id$.

\paragraph{The internal point of view}
\label{internal-groups}
By path induction, one can construct, for every two paths $p:a=b$ and $q:b=c$, a path $p\pcomp q : a=c$ called their \emph{concatenation}. Similarly, every path~$p:a=b$ admits an \emph{inverse} path $\sym p:b=a$. When~$A$ is a pointed groupoid, $\Loop A$ is a set, and these operations canonically equip this set with a group structure~\cite[Section~2.1]{hottbook}. A pointed groupoid thus provides an \emph{internal} notion of group in the sense that the group structure is induced by the inherent structure of the type (rather than being enforced by additional axioms). Moreover, this group structure only depends on the connected component of the canonical point, so that we might as well assume that the space is connected.

A \emph{delooping} of a group~$G$ is an internal representation of this group: it consists of a pointed connected groupoid $\B G$ together with an isomorphism of groups~$\Loop\B G\isoto G$. In such a situation, we often write $\Beq G:\Loop\B G=G$ for the identity induced under univalence by the isomorphism.
For instance, it is known that the circle is a delooping of~$\Z$: indeed, $\S1$ is a connected groupoid, and its fundamental group is~$\Z$~\cite[Section~8.1]{hottbook}.

\paragraph{Functoriality of delooping}
\label{delooping-functorial}
We now recall that deloopings are unique when they exist, which follows from the fact that we can always deloop group morphisms.
Suppose given two pointed connected groupoids~$A$ and~$B$: their deloopings $\Loop A$ and $\Loop B$ are groups.
Given a map $f:A\to B$, we have seen that there is an induced map $\Loop f:\Loop A\to\Loop B$, which is a group morphism. The loop space function thus induces a map
\[
  \Loop_{A,B}:(A\pto B)\to(\Loop A\toGroup\Loop B)
\]
and this map is always an equivalence. We provide a direct proof of this fact below, but this can also be recovered as the case $n=0$ of \cite[Corollary~12]{warn2023eilenberg} (see also~\cite[Lemma~6.5.1]{symmetry}).

\begin{proposition}
  \label{delooping-morphisms}
  Given pointed connected groupoids~$A$ and~$B$, the function $\Loop_{A,B}$ is an equivalence.
\end{proposition}
\begin{proof}
  By \cite[Theorem 4.4.3]{hottbook}, it is enough to show that the fiber of $\Loop_{A,B}$ at any morphism $g:\Loop A\toGroup\Loop B$ is contractible.
  By definition, the fiber of $\Loop_{A,B}$ at~$g$ is
  \[
    \fib{\Loop_{A,B}}g
    \qeq
    \Sigma(f:A\to B).\Sigma(\pt[f]:f\pt[A]=\pt[B]).(\Loop f=g)
  \]
  This type can be shown to be equivalent to the type
  \begin{equation}
    \label{eq:delooping-fun}
    \Sigma(f:A\to B).\Pi(a:A).C(a,f(a))
  \end{equation}
  which means that maps~$f:A\to B$ which are deloopable are those such that for every $a:A$ the image satisfies $C(a,f(a))$. Above, for $a:A$ and $b:B$, the type $C(a,b)$ is defined as
  \[
    \Sigma(\pi:(a=\pt[A])\to(b=\pt[B])).D(a,\pi)
  \]
  with $D(a,\pi)$ defined as
  \begin{equation}
    \label{eq:delooping-D}
    \Pi(p:a=\pt[A]).\Pi(q:\pt[A]=\pt[A]).\pi(p\pcomp q)=\pi(p)\pcomp g(q)
  \end{equation}
  see~\cite[Lemma~9]{warn2023eilenberg}. Namely, we have
  \begin{align*}
    \fib{\Loop_{A,B}}g
    &=\Sigma(f:A\to B).\Sigma(\pt[f]:f\pt[A]=\pt[B]).\Loop_{A,B}f=g
    \intertextright{by definition}
    &=\Sigma(f:A\to B).\Sigma(\pt[f]:f\pt[A]=\pt[B]).(q:\pt[A]=\pt[A])\to\Loop_{A,B}f(q)=g(q)
    \\
    \intertextright{by function extensionality}
    &=\Sigma(f:A\to B).\Sigma(\pt[f]:f\pt[A]=\pt[B]).(q:\pt[A]=\pt[A])\to\sym{\pt[f]}\pcomp\ap fq\pcomp\pt[f]=g(q)
    \\
    \intertextright{by definition of $\Loop_{A,B}$}
    &=\Sigma(f:A\to B).\Sigma(\pt[f]:f\pt[A]=\pt[B]).(q:\pt[A]=\pt[A])\to\ap fq\pcomp\pt[f]=\pt[f]\pcomp g(q)
    \\
    \intertextright{by identities on path composition}
    &=\Sigma(f:A\to B).\Sigma(\pi:(\Sigma(a:A).(a=\pt[A]))\to(f(a)=\pt[B])).\\
    &\qquad\qquad\Pi(q:\pt[A]=\pt[A]).\pi\pt[a]q=\pi\pt[A]\refl\pcomp g(q)
    \\
    \intertextright{by contractibility of singletons}
    &=\Sigma(f:A\to B).\Sigma(\pi:(a:A)\to(a=\pt[A])\to(f(a)=\pt[B])).\\
    &\qquad\qquad\Pi(q:\pt[A]=\pt[A]).\pi\pt[a]q=\pi\pt[A]\refl\pcomp g(q)
    \\
    \intertextright{by curryfication}
    &=\Sigma(f:A\to B).\Sigma(\pi:(a:A)\to(a=\pt[A])\to(f(a)=\pt[B])).\\
    &\qquad\qquad\Pi(q:\pt[A]=\pt[A]).\pi\pt[a](\refl\pcomp q)=\pi\pt[A]\refl\pcomp g(q)
    \\
    \intertextright{by unitality of $\refl$}
    &=\Sigma(f:A\to B).\Sigma(\pi:(a:A)\to(a=\pt[A])\to(f(a)=\pt[B])).\\
    &\qquad\qquad\Pi((a,p):\Sigma(a:A).a=\pt[A]).\Pi(q:\pt[A]=\pt[A]).\pi\,a\,(p\pcomp q)=\pi\,a\,(p)\pcomp g(q)
    \\
    \intertextright{by contractibility of singletons}
    &=\Sigma(f:A\to B).\Sigma(\pi:(a:A)\to(a=\pt[A])\to(f(a)=\pt[B])).\\
    &\qquad\qquad\Pi(a:A).\Pi(p:a=\pt[A]).\Pi(q:\pt[A]=\pt[A]).\pi\,a\,(p\pcomp q)=\pi\,a\,(p)\pcomp g(q)
    \\
    \intertextright{by curryfication}
    &=\Sigma(f:A\to B).\Pi(a:A).\Sigma(\pi:(a=\pt[A])\to(f(a)=\pt[B])).\Pi(p:a=\pt[A]).\\
    &\qquad\qquad\Pi(q:\pt[A]=\pt[A]).\pi(p\pcomp q)=\pi(p)\pcomp g(q)
    \\
    \intertextright{by type theoretic choice}
    &=\Sigma(f:A\to B).\Pi(a:A).C(a,f(a))
  \end{align*}
  Above, the ``type theoretic choice'' is the equivalence, for any types $A,B:\U$ and type family $P:A\to B\to\U$,
  \[
    \Pi(x:A).\Sigma(y:B).P\,x\,y
    \qquad\simeq\qquad
    \Sigma(f:A\to B).\Pi(x:A).P\,x\,(f\,x)
  \]
  see~\cite[Section~1.6]{hottbook}, and the ``contractibility of singletons'' is the fact that, for any type $A$, the type $\Sigma(x:A).(x=\pt)$ is contractible, see~\cite[Lemma 3.11.8]{hottbook}.
  %
  %
  By type theoretic choice again, we deduce that $\fib{\Loop_{A,B}}g$ is also equivalent to
  \begin{equation}
    \label{eq:delooping-sigma}
    \Pi(a:A).\Sigma(b:B).C(a,b)
  \end{equation}
  We show below that we have, for any $b:B$,
  \begin{equation}
    \label{eq:delooping-Cpt}
    C(\pt[A],b)\qeq(b=\pt[B])
  \end{equation}
  From there we deduce that $\Sigma(b:B).C(\pt[A],b)$ is equivalent to $\Sigma(b:B).(b=\pt[B])$ and is thus contractible (by contractibility of singletons). By path induction, we thus have
  \[
    \Pi(a:A).(a=\pt[A])\to\isContr(\Sigma(b:B).C(a,b))
  \]
  and thus
  \[
    \Pi(a:A).\ptrunc{a=\pt[A]}\to\isContr(\Sigma(b:B).C(a,b))
  \]
  because being contractible is a proposition. Since $A$ is connected, we deduce that the type $\Sigma(b:B).C(a,b)$ is contractible for every $a:A$, and the type \cref{eq:delooping-sigma} is thus also contractible. Therefore $\fib{\Loop_{A,B}}g$ is contractible, which is what we wanted to show.

  We are left with showing \cref{eq:delooping-Cpt}. It can be shown that, for any suitably typed function~$\pi$, the type $D(\pt[A],\pi)$ is equivalent to
  \[
    \Pi(q:\pt[A]=\pt[A]).\pi(q)=\pi(\refl)\pcomp g(q)
  \]
  Namely, the former implies the latter as a particular case and, conversely, assuming the second one, we have for $p:\pt[A]=\pt[A]$ and $q:\pt[A]=\pt[A]$,
  \[
    \pi(p\pcomp q)
    =
    \pi(\refl)\pcomp g(p\pcomp q)
    =
    \pi(\refl)\pcomp g(p)\pcomp g(q)
    =
    \pi(p)\pcomp g(q)
  \]
  because~$g$ preserves composition. From this, \cref{eq:delooping-Cpt} follows easily, \ie
  \[
    \Sigma(\pi:(\pt[A]=\pt[A])\to(b=\pt[B])).\pi(q)=\pi(\refl)\pcomp g(q)
  \]
  is equivalent to $b=\pt[B]$, since in the above type, the second component expresses that the function $\pi$ in the first component is uniquely determined by $\pi(\refl)$, which is an element of~$b=\pt[B]$.
\end{proof}

\noindent
Given a function $g:\Loop A\to\Loop B$, the previous proposition ensures that there is a unique function $f:A\pto B$ such that $\Loop f=g$. This function is noted $\B g$ and called the \emph{delooping} of~$g$.
%
%
As another immediate consequence of \cref{delooping-morphisms}, deloopings are unique:

\begin{proposition}\label{delooping-unique}
  Given two deloopings $\B G$ and $\B' G$ of a group~$G$, we have $\B G=\B' G$.
\end{proposition}
\begin{proof}
  Given two deloopings $\B G$ and $\B' G$, we have an equivalence $f:\Loop\B G\equivto\Loop\B' G$ with inverse $g$. These maps lift to morphisms $\B f:\B G\to\B' G$ and $\B g:\B' G\to\B G$. By the uniqueness statement in \cref{delooping-morphisms}, delooping of morphisms preserves composition and identities, so $\B f$ and $\B g$ form an equivalence.
\end{proof}

\noindent
Given a group morphism $f:G\to H$ such that both~$G$ and~$H$ admit deloopings (and we will see that this is actually always the case by \cref{PG-delooping,presented-delooping}), the \emph{delooping} of~$f$ is the morphism
\[
  \B f:\B G\to\B H
\]
associated, by \cref{delooping-morphisms}, to the morphism
$\transportinv{\Beq H}\circ f\circ\transport{\Beq G}:\Loop\B G\to\Loop\B
H$. By \cref{delooping-morphisms}, this operation is functorial in the sense
that it preserves identities and composition.

\paragraph{Equivalence between the two points of view}
Although this is not central in this article, we shall mention here the
fundamental equivalence provided by the above constructions; details can be
found in~\cite{symmetry}.
We write $\IGroup$ for the type of internal groups, \ie pointed connected
groupoids.


\begin{theorem}\label{internal-external-equivalence}
  The maps $\Loop:\IGroup\to\Group$ and $\B:\Group\to\IGroup$ form an
  equivalence of types.
\end{theorem}
\begin{proof}
  Given a group~$G$, we have $\Loop\B G=G$ by definition of~$\B G$. Given an
  internal group~$A$, we have $\B\Loop A\simeq A$ by \cref{delooping-unique}.
\end{proof}

\noindent
The above theorem thus states that the looping and delooping operators allow us to go
back and forth between the external and internal points of view of group
theory in homotopy type theory.

\paragraph{Internal group actions}
In a similar way as the traditional notion of group admits an internal
reformulation (\cref{internal-groups}), the notion of action also admits an
internal counterpart which can be defined as follows.
Given a group~$G$, an \emph{internal} action of $G$ on a set $A$ is a function
\[
  \alpha : \B G \to \Set
\]
such that $\alpha(\pt)=A$. Since $\Set$ is a
groupoid~\cite[Theorem 7.1.11]{hottbook}, by
\cref{internal-external-equivalence}, we have equivalences of types
\[
  (\B G\to\Set)
  \quad\equivto\quad
  (\Loop\B G\to\Loop(\Set,A))
  \quad\equivto\quad
  (G\to\Aut A)
\]
which shows that internal group actions correspond to external ones: the delooping operator internalizes an external group action, and the looping operator externalizes an internal group action.


\section{Delooping with torsors}
\label{torsor-delooping}
In this section, we recall a classical approach to constructing
deloopings of groups by using $G$-torsors, which originates in classical
constructions in algebraic topology~\cite{demazure1970groupes}. Most of the
material in this section is already known, which is why the proofs are not as
detailed. A more in-depth presentation can be found in recent works such
as~\cite{symmetry,buchholtz2023central}.

\paragraph{Group actions}
Given a group~$G$ and a set~$A$, an \emph{action} of $G$ on $A$ is a group
morphism from $G$ to $A \simeq A$, that is, a map $\alpha : G\to (A \simeq A)$
such that
\begin{align}
  \label{action-equations}
  \alpha(xy)&= \alpha(x) \circ \alpha(y)
  &
  \alpha(1) &= \id_A
\end{align}
for all $x,y:G$.

A \emph{$G$-set} is a set equipped with an action of~$G$, and we write $\GSet$
for the type of $G$-sets. We often simply denote a $G$-set by the associated
action $\alpha$ and write $\dom(\alpha)$ for the set on which~$G$ acts.

\begin{lemma}[\agda{GSetProperties}{isGroupoidGSet}]\label{gset-groupoid}
  The type $\GSet$ is a groupoid.
\end{lemma}
\begin{proof}
  The type of sets is a groupoid~\cite[Theorem~7.1.11]{hottbook}. Given a set
  $A$, the type of functions $A\to A$ is a set~\cite[Theorem 7.1.9]{hottbook}
  and thus a groupoid. Finally,
  the axioms~\cref{action-equations} of actions are propositions (because $A$
  is a set) and thus groupoids. We conclude since groupoids are closed under
  $\Sigma$-types~\cite[Theorem 7.1.8]{hottbook}.
\end{proof}

Given $G$-sets $\alpha$ and $\beta$, a \emph{morphism} between them consists of
a function $f:\dom{\alpha}\to \dom{\beta}$ which preserves the group action, in
the sense that for every $x:G$ and $a:\dom{\alpha}$, we have
\begin{equation}
  \label{G-set-morphism}
  \beta(x)(f(a))=f(\alpha(x)(a))
\end{equation}
A morphism which is also an equivalence is called an \emph{isomorphism} and we
write $\GSetEquiv\alpha\beta$ for the type of isomorphisms between $\alpha$ and
$\beta$. We write $\Aut(\alpha)$ for the type of automorphisms
$\GSetEquiv\alpha\alpha$, which is a group under composition.
%
The equalities between $G$-sets can be conveniently characterized as follows.

\begin{proposition}[\agda{GSetProperties}{GSet≡Decomp}]\label{Gset-equalities}
  Given two $G$-sets $\alpha$ and $\beta$, an equality between them consists of
  an equality $p:\dom{\alpha}=\dom{\beta}$ such that the function induced by
  transport along~$p$, namely $\transport p:\dom\alpha\to\dom\beta$, is a
  morphism of $G$-sets.
\end{proposition}
\begin{proof}
  The characterization of equalities between
  $\Sigma$-types~\cite[Theorem~2.7.2]{hottbook} entails that an equality between
  $(\dom{\alpha},\alpha)$ and $(\dom{\beta},\beta)$ is a pair of equalities
  \begin{align*}
    p:\dom{\alpha}&=\dom{\beta}
    &
    q:\transport{p}(\alpha)&=\beta 
  \end{align*}
  (we can forget about the equality between the components expressing the properties required for group actions since those are propositions). By~\cite[Lemma~2.9.6]{hottbook} and function extensionality, we finally have that the type of~$q$ is equivalent to the type $\beta(x)\circ\transport p=\transport p\circ\alpha(x)$.
\end{proof}

\noindent
It easily follows from this proposition that any equality between $G$-sets
induces an isomorphism of $G$-sets, as customary for equalities between
algebraic structures~\cite[Section~2.14]{hottbook}. In fact, this map from
equalities to isomorphisms can itself be shown to be an equivalence:

\begin{proposition}[\agda{GSetProperties}{GSetPath}]\label{GSet-SIP}
  Given $G$-sets $\alpha$ and $\beta$, the canonical function
  \[
    (\alpha=\beta)\to(\GSetEquiv\alpha\beta)
  \]
  is an equivalence.
  Moreover, given a $G$-set $\alpha$, the induced equivalence
  \[
    (\alpha=\alpha)\equivto(\GSetEquiv\alpha\alpha)
  \]
  is compatible with the canonical group structures on both types.
\end{proposition}
\begin{proof}

  This is actually an instance of a more general correspondence between
  equalities and isomorphisms of algebraic structures, which is known under the
  name of \emph{structure identity
    principle}, see~\cite{coquand2013isomorphism} and~\cite[Section~9.8]{hottbook}, and can be understood as a
  generalization of univalence for types having an algebraic structure.
\end{proof}

\paragraph{Connected components}
In the following, in order to define torsors, we will need to use a type theoretic counterpart for the notion of \emph{connected component} of a pointed type~$A$: this is the type of points which are merely connected to the distinguished point of~$A$. This type is noted~$\Comp A$ (or $\Comp(A,\pt)$ when we want to specify the distinguished element~$\pt$). Formally,
\[
  \Comp A
  \qdefd
  \Sigma(x:A).\ptrunc{\pt=x}
\]
This type is canonically pointed by $(\pt,\ptrunq{\refl})$.
This construction deserves its name because it produces a connected space, whose
geometry is the same as the original space around the distinguished point, as
shown in the following two lemmas.


\begin{lemma}[\agda{Comp}{isConnectedComp}]
  The type $\Comp A$ is connected.
\end{lemma}
\begin{proof}
  It can be shown that a type~$X$ is connected precisely when both $\ptrunc{X}$
  and $(x,y:X)\to\ptrunc{x=y}$ are inhabited, \ie when~$X$ merely has a point
  and any two points are merely equal~\cite[Exercise~7.6]{hottbook}.
  In our case, the type $\Comp A$ is pointed and thus $\ptrunc{\Comp A}$
  holds.
  Moreover, suppose that there are two points $(x,p)$ and $(y,q)$ in $\Comp A$
  with $x,y:A$, $p:\ptrunc{\pt=x}$ and $q:\ptrunc{\pt=y}$. Our goal is to
  show that $\ptrunc{(x,p)=(y,q)}$ holds, which is a proposition, so by
  elimination of propositional truncation, we can therefore assume that $p$
  (\resp $q$) has type $\pt=x$ (\resp $\pt=y$). Hence, we can construct a path
  $\sym p\pcomp q$ of type $x=y$, and therefore $(x,p)=(y,q)$ because the second
  components belong to a proposition by propositional truncation. We conclude
  that $\ptrunc{(x,p)=(y,q)}$ and finally that $\Comp A$ is connected.
\end{proof}

\begin{lemma}[\agda{Comp}{loopCompIsLoop}]
  \label{loop-comp}
  We have $\Loop\Comp A=\Loop A$.
\end{lemma}
\begin{proof}
  We begin by showing that the type
  \begin{equation}
    \label{eq:loop-comp-contr}
    \Sigma((x,t):\Comp A).(\pt=x)
  \end{equation}
  is contractible. In order to do so, observe that we have the following
  equivalence of types:
  \begin{align*}
    \Sigma((x,t):\Comp A).(\pt=x)
    &\simeq
    \Sigma((x,t):\Sigma(x:A).\ptrunc{\pt=x}).(\pt=x)
    \\
    &\equivto
    \Sigma(x:A).(\ptrunc{\pt=x})\times(\pt=x)
    \\
    &\equivto
    \Sigma((x,p):\Sigma(x:A).(\pt=x)).\ptrunc{\pt=x}
  \end{align*}
  using classical associativity and commutativity properties of
  $\Sigma$-types. Moreover, the type $\Sigma(x:A).(\pt=x)$ is
  contractible~\cite[Lemma~3.11.8]{hottbook}, therefore the whole type on the
  last line is a proposition (as a sum of propositions over a proposition), and
  therefore also the original type \cref{eq:loop-comp-contr}. We write $\pt'$
  for the element $(\pt,\ptrunc{\refl[\pt]})$ of $\Comp A$. The type
  \cref{eq:loop-comp-contr} is pointed by the canonical element
  $(\pt',\refl)$ and thus contractible as a pointed proposition.

  We have a morphism
  \[
    \begin{array}{r@{\ }c@{\ }c@{\ }c@{\ }c@{\ }l}
      F:&((x,t):\Comp A)&\to&(\pt'=(x,t))&\to&(\pt=x)\\
      &(x,t)&&p&\mapsto&\Cong\fst(p)
    \end{array}
  \]
  sending a path~$p$ to the path obtained by applying the first projection. It
  canonically induces a morphism
  \begin{align*}
    \Sigma((x,t):\Comp A).(\pt'=(x,t))&\to\Sigma((x,t):\Comp A).(\pt=x)\\
    ((x,t),p)&\mapsto((x,t),\Cong\fst(p))
  \end{align*}
  between the corresponding total spaces. Since the left member is contractible
  (by~\cite[Lemma~3.11.8]{hottbook} again) and the right member is also contractible
  (as shown above), this is an equivalence. By~\cite[Theorem~4.7.7]{hottbook},
  for every $x:\Comp A$, the fiber morphism $F x$ is also an equivalence. In
  particular, with $x$ being $\pt'$, we obtain $\Loop\Comp A\simeq\Loop A$ (as
  a type) and we can conclude by univalence. Note that the equivalence preserves
  the group structure so that the equality also holds in groups.
\end{proof}

\noindent
As a direct corollary of the two above lemmas, we have:

\begin{proposition}
  \label{B-loop}
  Given a pointed groupoid~$A$, $\Comp A$ is a delooping of~$\Loop A$.
\end{proposition}

\begin{remark}
  Some people write $\Aut A$ for $\Loop A$ and the above proposition states that
  we have $\B\Aut A=\Comp A$. For this reason, the (confusing) notation
  $\operatorname{BAut}A$ is also found in the literature for $\Comp A$.
\end{remark}


\paragraph{Torsors}
For any group~$G$, there is a canonical $G$-set called the \emph{principal $G$-torsor} and
noted~$\P G$, corresponding to the action of $G$ on itself by left
multiplication. Moreover, its group of automorphisms is precisely the group~$G$:

\begin{proposition}[\agda{Deloopings}{PGloops}]\label{loop-PG}
  Given a group~$G$, we have an equality of groups
  \[
    (\GSetEquiv{\P G}{\P G})=G
  \]
\end{proposition}
\begin{proof}
  The two functions
  \begin{align*}
    \phi:\Aut\P G&\to G
    &
    \psi:G&\to\Aut\P G
    \\
    f&\mapsto f(1)
    &
    x&\mapsto y\mapsto yx
  \end{align*}
  are group morphisms. Namely, given $f,g:\Aut\P G$, we have
  \begin{align*}
    \phi(g\circ f)&=g\circ f(1)=g(f(1)1)=f(1)g(1)=\phi(f)\phi(g)
    &
    \phi(\id)&=\id(1)=1
  \end{align*}
  and given $x,y:G$, we have for every $z:G$,
  \begin{align*}
    \psi(xy)(z)&=z(xy)=(zx)y=\psi(y)\circ\psi(x)(z)
    &
    \psi(1)(x)&=x1=\id(x)
  \end{align*}
  Moreover, they are mutually inverse. Namely, given $f:\Aut\P G$ and $x:G$, we
  have
  \begin{align*}
    \psi\circ\phi(f)(x)&=xf(1)=f(x1)=f(x)
    &
    \phi\circ\psi(x)&=1x=x
  \end{align*}
  We thus have $\Aut\P G\simeq G$ and we conclude by univalence.
\end{proof}

The type $\GSet$ is thus ``almost'' a delooping of~$G$. Namely, it is a groupoid (\cref{gset-groupoid}), it is pointed by $\P G$, and it satisfies~$\Loop\GSet = G$ by \cref{GSet-SIP,loop-PG}. It only lacks connectedness, which is easily addressed by restricting to the connected component.

\begin{definition}
  The type of \emph{$G$-torsors} is the connected component of $\P G$ in $\GSet$.
\end{definition}

\begin{theorem}[\agda{Deloopings}{torsorDeloops}]
  \label{PG-delooping}
  The type of $G$-torsors is a delooping of $G$.
\end{theorem}

\noindent
Note that the torsor construction only gives a delooping in a larger universe than the original group unless one makes additional assumptions such as the \emph{replacement axiom}~\cite[Axiom~18.1.8]{rijke2022introduction}.


\section{Generated torsors}
\label{generated-torsors}

\paragraph{Free groups}
Given a set~$X$, we write $\freegroup{X}$ for the \emph{free group} over~$X$~\cite[Theorem~6.11.6]{hottbook}. There is an inclusion function~$\incl:X\to\freegroup{X}$ which, by precomposition, induces an equivalence between morphisms of groups~$\freegroup{X}\toGroup G$ and functions~$X\to G$:
\[
  \begin{tikzcd}
    X\ar[d,"\incl"']\ar[r,"f"]&G\\
    \freegroup X\ar[ur,dotted,"\tilde f"']
  \end{tikzcd}
\]
We write $\freegroup{f}:\freegroup{X}\to G$ for the group morphism thus induced by a function~$f:X\to G$.
The elements of~$\freegroup X$ can be described as formal composites~$a_1\ldots a_n$ where each $a_i$ is an element of~$X$ or a formal inverse of an element of~$X$ (so that an element and an adjacent formal inverse cancel out).

\paragraph{Generated groups}
Fix a group~$G$. Given a set~$X$ and a map~$\gamma:X\to G$, we say that $X$ \emph{generates}~$G$ (with respect to~$\gamma$) when $\freegroup\gamma:\freegroup X\to G$ is surjective.
From now on, we suppose that we are in such a situation.
We now provide a variant of the construction of a delooping of~$G$ by $G$-torsors described in the previous section, taking advantage of the additional data of a generating set in order to obtain smaller and simpler constructions.
Note that here, contrarily to \cref{delooping-hit}, we only need a set of generators, not a full presentation.

\paragraph{Actions of sets} 
Given a type~$A$, we write $\End A$ for its type of \emph{endomorphisms}, \ie
maps $A\to A$. An \emph{action} of the set~$X$ on a set~$A$ is a morphism
$X\to\End A$, \ie a family of endomorphisms of~$A$ indexed by~$X$.
We write $\XSet$ for the type
\[
  \XSet
  \qdefd
  \Sigma(A:\Set).(X\to\End A)
  \]
of actions of~$X$. An element~$\alpha$ of this type consists in a set $\dom\alpha$ with a function
$\alpha:X\to\End(\dom\alpha)$ and is called an \emph{$X$-set}. A
\emph{morphism} between $X$-sets $\alpha$ and $\beta$ is a function
$f:\dom\alpha\to\dom\beta$ satisfying \cref{G-set-morphism} for every $x:X$.
The identities between $X$-sets can be characterized in a similar way as for
$G$-sets, see \cref{Gset-equalities}, and \cref{GSet-SIP} also extends in the
expected way.

Precomposition by $\gamma$ induces a function $U:\GSet\to\XSet$ which can be
thought of as a forgetful functor from $G$-sets to $X$-sets. Note that $U$
depends on~$\gamma$ but we leave it implicit for concision.

\paragraph{Applications of the generated delooping}
We have seen in the previous section that the connected
component of the principal $G$-torsor~$\P G$ in $G$-sets is a delooping
of~$G$. Our aim in this section is to show that this construction can be
simplified by taking the connected component of the restriction of $\P G$ to $X$-sets.

Before proving this theorem, which is formally stated as
\cref{generated-delooping} below, we shall first illustrate its use on a
concrete example. Consider $\Z_n$, the cyclic group with $n$ elements. We write
$s:\Z_n\to\Z_n$ for the successor (modulo $n$) function, which is an
isomorphism. By \cref{PG-delooping}, we know that the type
\[
  \Sigma(A:\GSet[\Z_n]).\ptrunc{\P{\Z_n}=A}
\]
of $\Z_n$-torsors is a model of~$\B\Z_n$.
This type is the connected component of the principal $\Z_n$-torsor $\P{\Z_n }$
in the universe~$\GSet[\Z_n]$ of sets with an action of $\Z_n$, \ie
sets~$A$ equipped with a morphism $ \alpha:\Z_n\to\Aut A $. Such a set~$A$
thus comes with one automorphism $\alpha(k)$ for every element~$k:\Z_n$,
therefore $k$ automorphisms in this case. However, most of them are superfluous:
$1$ generates all the elements of~$\Z_n$ by addition, so $\alpha(1)$
generates all the $\alpha(k)$ by composition because $\alpha(k)=\alpha(1)^k$.
The useful data of a $\Z_n$-set thus boils down to a set~$A$ together with one
automorphism $\alpha:\Aut A$ such that $\alpha^n=\id_A$.

Indeed, writing
$
  \SetEnd
  \defd
  \Sigma(A:\Set).\End A
$
for the type of all \emph{endomorphisms} (on any set), our theorem will imply
that the type
\begin{equation}
  \label{Zn-delooping}
  \Sigma((A,f):\SetEnd).\ptrunc{(\Z_n,s)=(A,f)}
\end{equation}
(the connected component of the successor modulo $n$ in the universe of set
endomorphisms) is still a delooping of $\Z_n$. Note that we didn't assume
that~$f$ is an isomorphism nor that it should verify $f^n=\id$. This is because
both properties follow from the fact that $f$ is in the connected component of
the successor (which satisfies those properties). Similarly, we do not need to
explicitly assume that the domain of the endomorphism is a set.

Our theorem thus allows to define, in a relatively simple way, types
corresponding to deloopings of groups. As recalled in the introduction, this is
particularly useful when one is not willing to use higher inductive types
(\eg because their definition, implementation and semantics are not entirely
mature). This is in fact the reason why this approach was used in UniMath to
define the circle~\cite{bezem2021construction}, and we provide a generic way to
similarly define other types.
We expect that it can be used to reason about groups and compute invariants such as their cohomology~\cite{cavallo2015synthetic,buchholtz2020cellular,brunerie2022synthetic}.
On a side note, one might be worried by the fact that we are ``biased'' (by
using a particular set of generators), which allows us to be more concise but
might make generic proofs more difficult compared to $G$-torsors: we expect that
this is not the case because in order to define the group~$G$ itself, one
usually needs to resort to a presentation, and thus is also biased in some sense...

\paragraph{The generated delooping}
In the following, we write $\P X$ for $U\P G$.

\begin{proposition}[\agda{XSetProperties}{theorem}]
  \label{PG-loop-restriction}
  We have a group equivalence
  $
  \Loop\P G\equivto\Loop \P X
  $.
\end{proposition}
\begin{proof}
  From \cref{Gset-equalities}, an element of $\Loop{\P G}$ consists of an equality
  $p:G=G$ in $\U$ such that
  \[
    \P G(g)\circ\transport p=\transport p\circ\P G(g)
  \]
  for every $g:G$. By function extensionality and the definition of the action $\P G$, this is equivalent to
  requiring, for every $g,z:G$ that
  \begin{equation}
    \label{eq:PG-condition}
    g(\transport p(z))=\transport p(gz)
  \end{equation}
  Note that the above equality is between elements of~$G$, which is a set, and is
  thus a proposition.
  Similarly, an element of $\Loop \P X$ consists of an equality $p:G=G$
  in~$\U$ satisfying
  \begin{equation}
    \label{eq:PX-condition}
    \gamma(x)(\transport p(z))=\transport p(\gamma(x)z)
  \end{equation}
  for every $x:X$ and $z:G$.

  Clearly, any equality $p:G=G$ in $\Loop\P G$ also belongs to $\Loop \P X$
  since the condition \cref{eq:PX-condition} is a particular case of
  \cref{eq:PG-condition}. We thus have a function
  $\phi:\Loop\P G\to\Loop \P X$.
  Conversely, consider an element $p:G=G$ of $\Loop \P X$, which has to satisfy
  \cref{eq:PX-condition} for every $x:X$ and $z:G$. Our aim is to show that it
  belongs to $\Loop\P G$. Given $g,z:G$, we thus want to show that
  \cref{eq:PG-condition} holds. Since~$\freegroup{\gamma}$ is surjective,
  because $X$ generates $G$, we know that there merely
  exists an element $u$ of~$\freegroup{X}$ such that
  $\freegroup{\gamma}(u)=g$. Since \cref{eq:PG-condition} is a proposition, by
  the elimination principle of propositional truncation, we can actually suppose
  given such a $u$, and we have
  \begin{align*}
    g(\transport p(y))
    &=\freegroup\gamma(u)(\transport p(y))&&\text{since $\freegroup{\gamma}(u)=g$}\\
    &=\transport p(\freegroup\gamma(u)y)&&\text{by repeated application of \cref{eq:PX-condition}}\\
    &=\transport p(gy)&&\text{since $\freegroup{\gamma}(u)=g$.}
  \end{align*}
  The second equality essentially corresponds to the commutation of the
  following diagram, where $u\defd x_1x_2\ldots x_n$ with $x_i:X$:
  \[
    \begin{tikzcd}
      G
      \ar[d,"\transport p"']
      \ar[r,"\gamma(x_1)"]
      &
      G
      \ar[d,"\transport p"']
      \ar[r,"\gamma(x_2)"]
      &
      G
      \ar[d,"\transport p"']
      \ar[r]
      &
      \ldots
      \ar[r]
      &
      G
      \ar[d,"\transport p"]
      \ar[r,"\gamma(x_n)"]
      &
      G
      \ar[d,"\transport p"]
      \\
      G
      \ar[r,"\gamma(x_1)"']
      &
      G
      \ar[r,"\gamma(x_2)"']
      &
      G
      \ar[r]
      &
      \ldots
      \ar[r]
      &
      G
      \ar[r,"\gamma(x_n)"']
      &
      G
    \end{tikzcd}
  \]
  This thus induces a function $\psi:\Loop \P X\to\Loop\P G$. The functions
  $\phi$ and $\psi$ clearly preserve the group structure (given by
  concatenation of paths) and are mutually inverse of each other, hence we have the equivalence we wanted.
\end{proof}

\begin{theorem}
  \label{generated-delooping}
  The type $\Comp \P X$ is a delooping of $G$.
\end{theorem}
\begin{proof}
  Since taking the connected component preserves loops spaces (\cref{B-loop}),
  we have that $\Loop\Comp\P X$ is equal to $\Loop\P X$, which in turn is equal
  to $\Loop\P G$ by \cref{PG-loop-restriction}, and thus to $G$ by
  \cref{PG-delooping}.
\end{proof}

The delooping of~$G$ constructed in the previous theorem is the component of~$\P X$
in $X$-sets:
\[
  \Sigma(A:\U).\Sigma(S:\isSet A).\Sigma(f:X\to\End A).\ptrunc{\P X=(A,S,f)}
\]
Since for any type~$A$, the type $\isSet A$ is a
proposition~\cite[Theorem~7.1.10]{hottbook}, and the underlying type of~$\P X$
is a set, the underlying type of any $X$-set in the connected component
of~$\P X$ will also be a set. As a consequence, the above type can be simplified slightly,
by dropping the requirement that~$A$ should be a set:

\begin{proposition}
  The type $\Sigma(A:\U).\Sigma(f:X \to\End A).\ptrunc{\P X=(A,f)}$ is a delooping
  of~$G$.
\end{proposition}

\noindent
For instance, the delooping \cref{Zn-delooping} of~$\Z_n$ can slightly be
simplified as
\[
  \Sigma((A,f):\Endomorphisms).\ptrunc{(\Z_n,s)=(A,f)}
\]
where $\Endomorphisms\defd\Sigma(A:\U).\End A$ is the type of all endomorphisms
in the universe.

\begin{example}
  \label{Dn}
  \Cref{generated-delooping} applies to every group for which a generating set
  is known (and, of course, the smaller the better). For instance, given a
  natural number~$n$, the \emph{dihedral group}~$D_n$ is the group of symmetries
  of a regular polygon with~$n$ sides. It has $2n$ elements and is generated by
  two elements $s$ (axial symmetry) and $r$ (rotation by an angle of
  $2\pi/n$). Hence the connected component of the symmetry and the rotation in
  the type of pairs of set endomorphisms, \ie
  \[
    \Sigma(A:\Set).\Sigma((f, g) : \End A\times\End A).\ptrunc{(D_n,s,r)=(A,f,g)}
  \]
  is a delooping of the dihedral group~$D_n$.
\end{example}

\paragraph{Alternative proof}
We would like to provide another proof of \cref{PG-loop-restriction}, which was suggested by an anonymous reviewer.
It is based on the idea that in order to show $\Loop\P X\equivto\Loop\P G$, it is enough to show that $U:\GSet\to\XSet$ is an \emph{embedding}, \ie that for every $\alpha,\beta:\GSet$ the induced function $U^=:(\alpha=\beta)\to(U\alpha=U\beta)$ between path spaces is an equivalence~\cite[Definition~4.6.1]{hottbook}.
It relies on the following results.

\begin{lemma}
  \label{fiberwise-embedding}
  Given a type~$A:\U$, type families $P,Q:A\to\U$ and $f:(a:A)\to P\,a\to Q\,a$, the map $\Sigma A.f:\Sigma A.P\to\Sigma A.Q$ is an embedding if and only if $f\,a:P\,a\to Q\,a$ is an embedding for every $a:A$.
\end{lemma}
\begin{proof}
  By definition, the map $\Sigma A.f$ is an embedding iff for every $(a,x)$ and $(a',x')$ in $\Sigma A.P$, the induced map
  \[
    (a, x) = (a', x') \to (a, f\,a\,x) = (a', f\,a'\,x')
  \]
  is an equivalence. By the characterization of equalities in $\Sigma$-types~\cite[Theorem~2.7.2]{hottbook}, this map corresponds to a map
  \[
    (\Sigma(p : a = a').\subst Pp(x) = x') \to (\Sigma(p : a = a').\subst Qp(f\,a\,x) = f\,a'\,x')
  \]
  By \cite[Theorem~4.7.7]{hottbook}, this is an equivalence if and only if the fiber map
  \[
    (\subst Pp(x) = x') \to (\subst Qp(f\,a\,x) = f\,a'\,x')
  \]
  is an equivalence for every $p : a = a'$. By path induction, this is true if and only if
  \[
    \ap{(f\,a)} : x = x' \to f\,a\,x = f\,a\,x'
  \]
  is an equivalence for all $a : A$, and $x,x' : P\,a$. By definition, this is the requirement that $f\,a$ is an embedding for all $a : A$.
\end{proof}

\begin{lemma}
  Given a morphism of groups $f:H\toGroup K$, we write $\GSet[f]:\GSet[K]\to\GSet[H]$ for the function induced by precomposition.
  If $f$ is surjective then $\GSet[f]$ is an embedding.
\end{lemma}
\begin{proof}
  Consider the map
  \[
    F:(A:\Set)\to(K\toGroup\Aut A)\to(H\toGroup\Aut A)
  \]
  obtained by precomposition by $f$. Since $f$ is surjective we have that $F\,A$ is an embedding for every set $A$~\cite[Lemma~10.1.4]{hottbook}. By \cref{fiberwise-embedding}, we deduce that $\Sigma\id_{\Set}.F$, which is $\GSet[f]$, is an embedding.
\end{proof}

\begin{lemma}
  The map $\GSet[\incl]:\GSet[\freegroup X]\to\GSet[X]$ is an embedding.
\end{lemma}
\begin{proof}
  By universal property of $\freegroup X$, given a set~$A$, the map
  \[
    (\freegroup X\toGroup\Aut A)\to(X\to\Aut A)
  \]
  obtained by precomposition by $\incl$ is an equivalence, and thus an
  embedding. Moreover, the property of being an isomorphism is a proposition, hence the forgetful map
  \[
    (X\to\Aut A)\to(X\to\End A)
  \]
  is an embedding because its fibers are propositions. Since embeddings are
  stable under composition, we deduce that the induced map
  \[
    F:(A:\Set)\to(\freegroup X\toGroup\Aut A)\to(X\to\End A)
  \]
  is such that $F\,A$ is an embedding for every set $A$. By
  \cref{fiberwise-embedding}, the map $\Sigma\id_{\Set}.F$, which is $\GSet[\incl]$, is
  thus an embedding.
\end{proof}

\begin{proposition}
  The map $U:\GSet\to\XSet$ is an embedding.
\end{proposition}
\begin{proof}
  Given a morphism of groups $f:H\to K$, we write $\GSet[f]:\GSet[K]\to\GSet[H]$ for the function induced by precomposition.
  In particular, by definition, we have $U\defd\GSet[\gamma]$.
  The function $\gamma:X\to G$ can be decomposed as $\freegroup\gamma\circ\incl$, and therefore $\GSet[\gamma]$ can be decomposed as $\GSet[\incl]\circ\GSet[\freegroup\gamma]$:
  \[
    \begin{tikzcd}
      &\ar[dl,"\freegroup\gamma"']\freegroup X\\
      G&&\ar[ll,"\gamma"]\ar[ul,"\incl"']X
    \end{tikzcd}
    \qquad\qquad\qquad
    \begin{tikzcd}
      &\GSet[\freegroup X]\ar[dr,"{\GSet[\incl]}"]&\\
      \GSet\ar[ur,"{\GSet[\freegroup\gamma]}"]\ar[rr,"U"']&&\XSet
    \end{tikzcd}
  \]
  By previous lemmas, both maps are embeddings so that $U$ is an embedding by composition.
\end{proof}


\section{Delooping using higher inductive types}
\label{delooping-hit}
We now recall an alternative construction of deloopings of groups, when higher inductive types are available in the theory. We then explain how to refine it when a presentation is known for the group.

\paragraph{Delooping as a higher inductive type}
Given a group~$G$, its delooping should have a point~$\pt$ and a loop for every element of the group. Moreover, we should ensure that the multiplication of $G$ coincides with concatenation on the loop space, and that the resulting type is a pointed connected groupoid. This suggests considering a higher inductive type, denoted $\K(G,1)$, with the following constructors
\[
  \begin{array}{l@{\ :\ }l}
    \pt&{\K(G,1)}\\
    \cloop&G\to\pt=\pt\\
    \cloopcomp&(x,y:G)\to\cloop x\pcomp\cloop y=\cloop(xy)\\
    \ctrunc&\isGroupoid(\K(G,1))
  \end{array}
\]
This construction was first proposed by Finster and Licata. They also showed, using the encode-decode method, that it is a delooping of the original group, \ie $\Loop\K(G,1)=G$, see~\cite[Theorem~3.2]{licata2014eilenberg}. Note that we only ask here that $\cloop$ preserves multiplication (with $\cloopcomp$), because it can be shown that this implies preservation of unit and inverses. In particular, preservation of unit renders superfluous one of the constructors present in the original definition~\cite{licata2014eilenberg}:

\begin{lemma}[\agda{EM}{loop-id}]
  In $\K(G,1)$, we have $\cloop 1=\refl$.
\end{lemma}
\begin{proof}
  Omitting associativity and unitality of path composition, we have
  \begin{align*}
    \cloop 1=\sym{(\cloop 1)}\cdot\cloop 1\cdot\cloop 1=\sym{(\cloop 1)}\cdot\cloop 1=\refl
  \end{align*}
  where the equality in the middle is derived from~$\cloopcomp 1$.
\end{proof}

In the following, we define a variant of this higher inductive type for presented groups, which is smaller and gives rise to computations closer to traditional group theory.

\paragraph{Presentations of groups}
Any free group~$\freegroup X$ on a set~$X$ admits a delooping as a wedge of an $X$-indexed family of circles. The corresponding type~$\bigvee_X\S1$ can be described as the coequalizer
\begin{equation}
  \label{wedge-of-circles}
  \begin{tikzcd}
    X\ar[r,shift left]\ar[r,shift right]&1\ar[r,dotted]&\bigvee_X\S1
  \end{tikzcd}
\end{equation}
or, equivalently, as the higher inductive type
\[
  \begin{array}{l@{\ :\ }l}
    \pt&\bigvee_X\S1\\
    \cloop&X\to\pt=\pt
  \end{array}
\]
generated by the two constructors $\pt:\bigvee_X\S1$ and $\cloop:X\to\pt=\pt$.

\begin{proposition}
  \label{bouquet-loop-space}
  We have $\Loop\bigvee_X\S1=\freegroup{X}$, \ie the above type is a $\B\freegroup X$.
\end{proposition}
\begin{proof}
  The fact that $\bigvee_X\S1$ is a delooping of~$\freegroup X$ is not too difficult to show when~$X$ has decidable equality, see~\cite[Exercise 8.2]{hottbook} and~\cite{kraus2018free}, but the general case is more involved and was recently proved in~\cite{warn2023path}: the main issue is to show that this type is a groupoid.
\end{proof}

A group \emph{presentation} $\pres XR$ consists of a set~$X$ of
\emph{generators}, a set~$R$ of \emph{relations}, and two functions
$\fst,\snd:R\to\freegroup X$ respectively associating to a relation its
\emph{source} and \emph{target}. We often write $r:u\To v$ for a
relation~$r$ with $u$ as source and~$v$ as target.
Given such a presentation~$P$, the corresponding \emph{presented} group~$\pgroup P$ is the set quotient $\freegroup{X}/R$ of the free group on $X$ under the
smallest congruence identifying the source and the target of every
relation~$r:R$. This type can be described as the type
$\pgroup P\defd\strunc{{\freegroup X}\hq R}$ obtained by taking the set
truncation of the coequalizer
\[
  \begin{tikzcd}
    R
    \ar[r,shift left,"\fst"]
    \ar[r,shift right,"\snd"']
    &
    \freegroup X
    \ar[r,dotted,"\kappa"]
    &
    \freegroup X\hq R
  \end{tikzcd}
\]
From this also follows a description of~$\pgroup P$ as a higher inductive type:
\[
  \begin{array}{l@{\ :\ }l}
    \constructor{word}&\freegroup X\to\pgroup P\\
    \constructor{rel}&(r:R)\to\constructor{word}(\fst(r))=\constructor{word}(\snd(r))\\
    \ctrunc&\isSet(\pgroup P)
  \end{array}
\]
Note that when~$R$ is the empty type, the presented group is the free group on the generators.

\paragraph{A smaller delooping}
\label{delooping-presented-hit}
Suppose given a group~$G$ along with a presentation $P\defd\pres XR$, \ie such that $G=\pgroup P$. We define the type $\B P$ as the following higher inductive type:
\[
  \begin{array}{l@{\ :\ }l}
    \pt&{\B P}\\
    \cgen&X\to(\pt=\pt)\\
    \crel&(r:R)\to(\freegroup\cgen(\fst(r))=\freegroup\cgen(\snd(r)))\\
    \ctrunc&\isGroupoid(\B P)
  \end{array}
\]
This type is generated by a point~$\pt$, then the constructor~$\cgen$ adds a
loop~$\ul a:\pt=\pt$ for every generator~$a$, the constructor~$\crel$ adds
an equality
$\ul a_1\pcomp\ul a_2\pcomp\ldots\pcomp\ul a_n=\ul b_1\pcomp\ul
b_2\pcomp\ldots\pcomp\ul b_m$ for each relation
$a_1\ldots a_n\To b_1\ldots b_m$, and the constructor~$\ctrunc$ formally takes
the groupoid truncation of the resulting type. Note that, because of the
presence of $\freegroup\cgen:\freegroup X\to(\pt=\pt)$ in the type
of~$\crel$, the above inductive type is not accepted as is in standard proof
assistants such as Agda. However, a definition can be done in two stages, by
first considering $\bigvee_X\S1$ (\ie the type generated only by $\pt$ and $\cgen$), and then defining a
second inductive type further quotienting this type (\ie adding the
constructors $\crel$ and $\ctrunc$), see \agda{EM}{Delooping}. Also, the
definition of $\freegroup\cgen$ requires the group structure on $\bigvee_X\S1$: the group operations are easily defined from operations on paths (reflexivity, concatenation, symmetry), but the fact that it is a groupoid is non-trivial (see \cref{bouquet-loop-space}).
Our main result in this section is the following:

\begin{theorem}[\agda{EM}{theorem}]
  \label{presented-delooping}
  Given a presentation~$P\defd\pres XR$, the type $\B P$ is a delooping of the group~$\pgroup P$.
\end{theorem}
\begin{proof}
  %
  By induction on~$\B P$, we can define a function $f:\B P\to\K(\pgroup P,1)$ such that
  $f\pt\defd\pt$, and $\ap f(\cgen a)\defd\cloop\pgroup a$ for all $a:X$. It can
  be shown that $f$ is then such that
  $\Cong f(\freegroup\cgen u)=\cloop\pgroup u$, for any $u:\freegroup X$. We can
  therefore define the image $\Cong f(r)$ on a relation $r:u\To v$ as the
  composite of equalities
  \[
    \Cong f(\freegroup\cgen u)=\cloop\pgroup u=\cloop\pgroup v=\Cong f(\freegroup\cgen v)
  \]
  where the equality in the middle follows from the fact that
  $\pgroup u=\pgroup v$ because of the relation~$r$.

  In the other direction, the group morphism $\freegroup\cgen:\freegroup{X}\to \Loop\B P$ preserves relations (by~$\crel$), and thus induces a quotient morphism $g':\pgroup P\to\Loop\B P$.
  We can thus consider the function $g:\K(\pgroup P,1)\to\B P$ such that $g(\pt)=\pt$, for $x:\pgroup P$ we have $\ap g(\cloop x)=g'(x)$, and for $x,y:\pgroup P$ the image of $\cloopcomp x\,y$ is canonically induced by the fact that $g'$ preserves group multiplication.

  Since $\K(\pgroup P,1)$ is a groupoid, in order to show that $f(g(x))=x$ for every $x:\K(\pgroup P,1)$, it is enough to show that it holds for $x\defd\pt$, which is the case by definition of $f$ and $g$, and that this property is preserved under $\cloop x$ for $x:\pgroup P$. This follows from the fact that we have $\ap f(g'(x))=\cloop x$ for any~$x:\pgroup P$ (this is easily shown by induction on~$x$).
  Conversely, we have to show that $g(f(x))=x$ holds for $x:\B P$. Again, this is shown by induction on~$x$.
  %
  %
\end{proof}

\noindent
Interestingly, the careful reader will note that the fact that the types $X$ and $R$ are sets does not play a role in the proof: in fact, those assumptions can be dropped here.
Also, note that we do not need the choice of a representative in $\freegroup X$ for every element of $\pgroup P$ in order to define the function $g$ from $\freegroup\cgen$ in the above proof: intuitively, this is because the induced function~$g$ does not depend on such a choice of representatives.
Finally, we should mention here that a similar result is mentioned as an exercise in \cite[Example 8.7.17]{hottbook}; the proof suggested there is more involved since it is based on a generalized van Kampen theorem.

\begin{example}
  The dihedral group $D_5$, see \cref{Dn}, admits the presentation
  \[
    \pres{r,s}{r^4=srs,sr^2s=r^3,rsr=s,r^3s=sr^2,sr^3=r^2s,s^2=1}
  \]
  Hence, by \cref{presented-delooping} we can construct a delooping of $D_5$ as an
  higher inductive type generated by two loops (corresponding to~$r$ and $s$)
  and six $2$-dimensional cells (corresponding to the relations). Note that this
  is much smaller than $\K(D_5,1)$ (it has 2 instead of 10 generating loops, and
  6 instead of 100 relations), thus resulting in shorter proofs when reasoning
  by induction.
\end{example}

\begin{example}
  Any group~$G$ admits a presentation, the \emph{standard presentation}, with one generator $\ul a$ for every element $a:G$, and relations $\ul a\,\,\ul b=\ul{ab}$ for every pair of generators, as well as $\ul 1=1$. By applying \cref{presented-delooping}, we actually recover the inductive type~$\K(G,1)$ as delooping of~$G$.
\end{example}


\section{2-polygraphs}
\label{polygraphs}
In \cref{delooping-hit}, we constructed the higher inductive type $\B P$, which is smaller than the standard construction of Finster and Licata~\cite{licata2014eilenberg} for $\B G$ when a presentation~$P$ for~$G$ is known. In this section, we introduce the notion of a \emph{2-polygraph}, which internalizes the data necessary to describe such higher inductive types and manipulate them. This notion, which originates in the study of strict higher categories~\cite{burroni1993higher,polygraphs}, has recently been ported to homotopy type theory by Kraus and von Raumer~\cite{kraus2022rewriting} and further studied in~\cite{coh-tietze}. Here, we introduce the notions of generated and presented types of a 2-polygraph, so that the type presented by a 2-polygraph~$P$ corresponding to a presentation of a group~$G$ is precisely $\B P$. We also illustrate the usefulness of 2-polygraphs by introducing Tietze transformations, which allow computations to be performed on them while preserving the presented type (\cref{tietze-invariant}). These operations can thus be used to transform a presentation into one more amenable to computation (such as a convergent one, see \cref{knuth-bendix}).

\paragraph{Polygraphs}
A \emph{0-polygraph}~$P$ is a set $P_0$. A \emph{0-sphere} in such a polygraph is a pair of elements of $P_0$, and we write
$
\Spheres0 P
\defd
P_0\times P_0
$
for the type of spheres in~$P$.

A \emph{1-polygraph}~$P$ consists of a 0-polygraph~$P_0$ together with a function $P_1:\Spheres0P\to\Set$ which to a sphere $(x,y)$ associates the set $P_1(x,y)$ of \emph{1-generators} from~$x$ to~$y$ ($x$ and $y$ are sometimes respectively referred to as the \emph{source} and \emph{target} of the 1-generator). We write $\freegpd P_1:\Spheres0P\to\Set$ for the function which to a 0-sphere $(x,y)$ associates the set $\freegpd P_1(x,y)$ of \emph{1-cells} from~$x$ to~$y$. Those consist of formal composable zig-zags of 1-generators, \ie formal composites of generators or inverses of those, which can be formally defined as sequences
\[
  a_1^{\epsilon_1}\cdot a_2^{\epsilon_2}\cdots a_n^{\epsilon_n}
\]
where $a_i^{\epsilon_i}$ is either of the form $a_i^+$ with $a_i:P_1(x_{i-1},x_i)$ or $a_i^-$ with $a_i:P_1(x_i,x_{i-1})$ for some sequence of 0-generators $x_0,x_1,\ldots,x_n:P_0$, with $x_0=x$ as \emph{source} and $x_n=y$ as \emph{target}. Given a 0-cell~$x$, we write $\id_x:\freegpd P_1(x,x)$ for the nullary composition (the trivial 1-cell). Given two 1-cells $u:\freegpd P_1(x,y)$ and $v:\freegpd P_1(y,z)$, we write $u\comp0 v:\freegpd P_1(x,z)$ for the 1-cell obtained by concatenation of the two lists.
Note that our definition of 1-cells is not ``coherent'' in the sense that we would expect relations such as $a^-\cdot a^+=\id$ to hold for any 1-generator~$a$, as well as associated higher coherences. However, we will see that those can be introduced when needed, as explicit 2-cells in the definition of $\freegpd P_2$ below. 
A \emph{1-sphere} in $P$ is a pair of two 1-cells with the same source and target, and we write
\[
  \Spheres1 P
  \quad\defd\quad
  \Sigma((x,y):\Spheres0 P).(\freegpd P_1(x,y)\times\freegpd P_1(x,y))
\]
for the type of 1-spheres in~$P$. We often simply write $(u,v)$ instead of $((x,y),(u,v))$ for a 1-sphere.

\newcommand{\pred}[1]{#1'}

A \emph{2-polygraph}~$P$ consists of a 1-polygraph~$\pred P=(P_0,P_1)$ as above, together with a function $P_2:\Spheres1 P\to\Set$ which to a 1-sphere $(u,v)$ associates the set $P_2(u,v)$ of 2-generators of the 2-polygraph.

\begin{remark}
  \label{fibered-polygraphs}
  The definition of polygraphs is ``indexed'' in the sense that we provide sets of $(n{+}1)$\nbd-gene\-rators which are indexed by $n$-spheres. By the Grothendieck duality (see \cref{grothendieck-duality}), we can also provide an equivalent fibered definition, where we provide the (total) set of all $(n{+}1)$-generators along with a function associating their boundary $n$-sphere.

  More explicitly, a \emph{fibered 1-polygraph} consists of a 0-polygraph $P_0$ together with a set~$P_1$ and a function $\partial:P_1\to\Spheres0 P$ which to every 1-cell associates its boundary 0-sphere. Similarly, a \emph{fibered 2-polygraph} consists in a fibered 1-polygraph together with a set $P_2$ and a function $\partial:P_2\to\Spheres1 P$. In this context, we often write $a:x\to y$ in order to indicate that $a$ is an element of~$P_1$ with $\partial(a)=(x,y)$, and similarly we write $\alpha:u\To v$ for $\alpha:P_2$ with $\partial\alpha=(u,v)$.

 For instance, any (non-fibered) 1-polygraph $P$ induces a fibered 1-polygraph $Q$ with
  \begin{align*}
    Q_0&\eqdef P_0
    &
    Q_1&\defd\Sigma((x,y):\Spheres0 P).P_1(x,y) 
  \end{align*}
  and $\partial:Q_1\to \Spheres0 P$ is the first projection. Conversely, any fibered 1-polygraph~$Q$ induces a non-fibered one~$P$ by setting $P_0\eqdef Q_0$ and $P_1(x,y)=\fib\partial(x,y)$. These two operations are mutually inverse, and a similar equivalence can be constructed for 2-polygraphs.
\end{remark}

\paragraph{Generated type}
\newcommand{\gtype}{\overline}
Given a $1$-polygraph~$P$, the type \emph{generated} by~$P$ is the type $\gtype{P}$ described as the following higher inductive type
\[
  \begin{array}{l@{\ :\ }l}
    \cpt&{P_0\to\gtype P}\\
    \cgen&((x,y):\Spheres0P)\to(a:P_1(x,y))\to(\cpt x=\cpt y)
  \end{array}
\]
We write
\[
  \freegpd\cgen:((x,y):\Spheres0P)\to(u:\freegpd P_1(x,y))\to(\cpt x=\cpt y)
\]
for the expected extension of $\cgen$ to cells, defined by
\begin{align*}
  \freegpd\cgen(u\cdot v)&=\freegpd\cgen(u)\pcomp\freegpd\cgen(v)
  &
  \freegpd\cgen(a^+)&=\cgen a
  \\
  \freegpd\cgen(\id)&=\refl
  &
  \freegpd\cgen(a^-)&=\sym{(\cgen a)}
\end{align*}
More abstractly, $\gtype P$ can be computed as the following coequalizer
\[
  \begin{tikzcd}[column sep=huge]
    \Sigma\Spheres0P.P_1\ar[r,shift left,"\fst\fst"]\ar[r,shift right,,"\snd\fst"']&P_0\ar[r,dotted]&\gtype P
  \end{tikzcd}
\]
where the two maps $\Sigma\Spheres0P.P_1\to P_0$ respectively send a pair $((x,y),a)$ to $x$ and $y$: given a 1-generator, we identify its source and target in $\gtype P$.

Given a 2-polygraph~$P$, the type $\gtype P$ \emph{generated} by~$P$ is the type generated by the higher inductive type
\[
  \begin{array}{l@{\ :\ }l}
    \cpred&{\gtype{\pred P}\to\gtype P}\\
    \crel&((u,v):\Spheres1P)\to(\alpha:P_2(u,v))\to(\freegpd\cgen(u)=\freegpd\cgen(v))
  \end{array}
\]
Note that, by the first constructor, we have a canonical inclusion in~$\gtype P$ of the type generated (in the above sense) by the underlying 1-polygraph~$P'$.

\bigskip

In order to provide an alternative definition of the type generated by a polygraph as a colimit, we first need to introduce some notations.
We write~$I$ for the \emph{interval} type, generated by two points and a path between them. This type is of course contractible, so that we could use the equivalent type~$1$ instead, but we believe that its use clarifies notations and intuitions. We also write $\G$ for the \emph{globe} type, which is defined as the pushout on the left and corresponds to the space pictured on the right
\begin{align}
  \label{globe-pushout}
  \begin{tikzcd}[ampersand replacement=\&]
    2 \& I\\
    I \& \G
    \arrow[from=1-1, to=1-2,"i"]
    \arrow[from=1-1, to=2-1,"i"']
    \arrow[from=1-2, to=2-2]
    \arrow[from=2-1, to=2-2]
    \arrow["\lrcorner"{anchor=center, pos=0.125, rotate=180}, draw=none, from=2-2, to=1-1]
  \end{tikzcd}
  &&
  \G\defd
  \begin{tikzcd}[ampersand replacement=\&]
    \cdot\ar[r,-,bend left=50,shorten=-4pt]\ar[r,-,bend right=50,shorten=-4pt]\&\cdot
  \end{tikzcd}
\end{align}
Above, the map $i:2\to I$ is the canonical inclusion (sending the elements of $2$ to the endpoints of~$I$). One should think of $\G$ as a two intervals glued on their endpoints, thus forming a circle: these intervals are sometimes respectively referred to as the northern and southern hemispheres. It is easily shown that~$\G$ equivalent to $\S{1}$. By the universal property of the pushout, any two maps $f,g:I\to X$ together with an identification $f\circ i=g\circ i$ induces a map denoted $f+g:\G\to X$.

With those notations, the type~$\gtype P$ generated by a 2-polygraph~$P$ can be described as the pushout
\begin{equation}
  \label{2pol-gtype-pushout}
  \begin{tikzcd}
    \G\times\Sigma\Spheres1P.P_2\ar[d]\ar[r,"\snd"]\ar[dr,phantom,pos=1,"\ulcorner"]&\Sigma\Spheres1P.P_2\ar[d]\\
    \gtype{P'}\ar[r]&\gtype{P}
  \end{tikzcd}  
\end{equation}
where $P'$ is the underlying 1-polygraph of~$P$, the topmost horizontal map is the second projection and the leftmost vertical map is obtained by currying from the map
\begin{align*}
  \Sigma\Spheres1P.P_2&\to(O\to\gtype{P'})\\
  (((x,y),(u,v)),\alpha)&\to\freegpd\cgen u+\freegpd\cgen v
\end{align*}
sending a 2-generator to the globe in $\gtype{P'}$ corresponding to its boundary.
This construction of~$\gtype{P}$ can be summarized in the following diagram:
%
\begin{equation}
  \label{polygraph-colimit}
  \begin{tikzcd}
     &&\G\times\Sigma\Spheres1P.P_2\ar[d]\ar[r,"\snd"]\ar[dr,phantom,pos=1,"\ulcorner"]&\Sigma\Spheres1P.P_2\ar[d]\\
     \Sigma\Spheres0P.P_1\ar[r,shift left,"\fst\fst"]\ar[r,shift right,,"\snd\fst"']&P_0\ar[r]&\gtype{P'}\ar[r]&\gtype{P}
  \end{tikzcd}
\end{equation}
Note that this is an instance of a cellular type in the sense of~\cite{buchholtz2020cellular}.

\paragraph{Presented type}
\newcommand{\ptype}[1]{\gtrunc{\gtype #1}}
The type \emph{presented} by a 2-polygraph~$P$ is $\gtrunc{\gtype P}$.

\bigskip

When $P_0\eqdef 1$ contains only one element, a (fibered) 2-polygraph~$P$ encodes precisely the data of a presentation $\pres XR$ of a group~$G$ with $X\eqdef P_1$ as generators and $R\eqdef P_2$ as relations, each element $\alpha:P_2$ with $\partial(\alpha)=(u,v)$ encoding a relation identifying $u$ and $v$. The type presented by this polygraph is then precisely the one we described in \cref{delooping-presented-hit}. Conversely, the notion of 2-polygraph can be thought of as a mild generalization of the notion of group presentation, where we allow multiple 0-cells.

\begin{proposition}
  \label{pres-pol}
  Given a group~$G$ with a presentation $\pres XR$, consider the fibered 2\nbd-poly\-graph $P$ with $P_0\defd 1$, $P_1\defd X$, $P_2\defd R$ and $\partial(u,v)=(u,v)$. The type presented by~$P$ is equal to $\B\pres XR$ (in the sense of \cref{presented-delooping}).
\end{proposition}

By analogy with $\freegpd P_1$, we can define a function $\freegpd P_2:\Spheres1 P\to\Set$, such that the elements of~$\freegpd P_2(u,v)$ are called the \emph{2-cells} (or \emph{relations}) from~$u$ to~$v$ in~$P$. It is defined as the smallest family of sets $\freegpd P_2:\Spheres1 P\to\Set$ such that
\begin{itemize}
\item for any 1-sphere $(u,v)$, every 2-generator $\alpha:P_2(u,v)$ induces an element in $\freegpd P_2(u,v)$, still noted~$\alpha$,
  \[
    \begin{tikzcd}
      x
      \ar[r,bend left=60,"u",""'{name=U,pos=0.5}]
      \ar[r,bend right=60,"v"'{pos=0.53},""{name=V,pos=0.53}]
      &
      y
      \ar[from=U,to=V,Rightarrow,"\alpha"']
    \end{tikzcd}
  \]
\item for any 1-spheres $((x,y),(u,u'))$ and $((y,z),(v,v'))$, any 2-cells $\alpha:\freegpd P_2(u,u')$ and $\beta:\freegpd P_2(v,v')$ induce a 2-cell $\alpha\comp0\beta:\freegpd P_2(u\comp0 v,u'\comp0 v')$,
  \[
    \begin{tikzcd}[cells={nodes={minimum size=5mm}}]
      x
      \ar[r,bend left=60,"u",""'{name=U}]
      \ar[r,bend right=60,"u'"',""{name=U',pos=.53}]
      &
      y
      \ar[r,bend left=60,"v",""'{name=V}]
      \ar[r,bend right=60,"v'"'{pos=.49},""{name=V',pos=.47}]
      &
      z
      \ar[from=U,to=U',Rightarrow,"\alpha"']
      \ar[from=V,to=V',Rightarrow,"\beta"'{pos=.6}]
    \end{tikzcd}
    \qquad\rightsquigarrow\quad
    \begin{tikzcd}
      x
      \ar[rr,bend left=40,"u\comp0 v",""'{name=UV}]
      \ar[rr,bend right=40,"u'\comp0 v'"'{pos=.51},""{name=UV',pos=.5}]
      &&
      z
      \ar[from=UV,to=UV',Rightarrow,"\alpha\comp0\beta"']
    \end{tikzcd}
  \]
\item for any 1-spheres $(u,v)$ and $(v,w)$, any 2-cells $\alpha:\freegpd P_2(u,v)$ and $\beta:\freegpd P_2(v,w)$ induce a 2-cell $\alpha\comp1 \beta:\freegpd P_2(u,w)$,
  \[
    \begin{tikzcd}
      x
      \ar[r,bend left=60,"u"]
      \ar[r,"v"description]
      \ar[r,bend right=60,"w"'{pos=.53}]
      \ar[r,phantom,bend left=30,"\Da\alpha"]
      \ar[r,phantom,bend right=30,"\Da\beta"]
      &y
    \end{tikzcd}
    \qquad\rightsquigarrow\quad
    \begin{tikzcd}[column sep=large]
      x
      \ar[r,bend left=60,"u",""'{name=U}]
      \ar[r,bend right=60,"w"',""{name=W,pos=.52}]
      &y
      \ar[from=U,to=W,Rightarrow,"\alpha\comp1\beta"']
    \end{tikzcd}
  \]
\item any 1-cell $u:\freegpd P_1(x,y)$ induces a 1-cell $\id_u:\freegpd P_2(u,u)$,
  \[
    \begin{tikzcd}[cells={nodes={minimum size=5mm}}]
      x
      \ar[r,bend left=60,"u",""'{name=U}]
      \ar[r,bend right=60,"u"'{pos=.53},""{name=U',pos=.53}]
      &y
      \ar[from=U,to=U',Rightarrow,"\id_u"']
    \end{tikzcd}
  \]
\item for any 1-sphere $(u,v)$, any 2-cell $\alpha:\freegpd P_2(u,v)$ induces a 2-cell $\inv\alpha:\freegpd P_2(v,u)$,
  \[
    \begin{tikzcd}[cells={nodes={minimum size=5mm}}]
      x
      \ar[r,bend left=60,"u",""'{name=U}]
      \ar[r,bend right=60,"v"',""{name=V,pos=.53}]
      &y
      \ar[from=U,to=V,Rightarrow,"\alpha"'] 
    \end{tikzcd}
    \qquad\rightsquigarrow\quad
    \begin{tikzcd}[cells={nodes={minimum size=5mm}}]
      x
      \ar[r,bend left=60,"u",""'{name=U}]
      \ar[r,bend right=60,"v"',""{name=V,pos=.53}]
      &y
      \ar[from=V,to=U,Rightarrow,"\alpha^{\!{-}\!1}"]      
    \end{tikzcd}
  \]
\item for any 1-generator $a:P_1(x,y)$ induces a pair of 2-cells $\lambda_a:\freegpd P_2(a^-\cdot a^+,\id_y)$ and $\rho_a:\freegpd P_2(a^+\cdot a^-,\id_x)$:
  \[
    \begin{tikzcd}
      &x\ar[phantom,""'{name=X,below=5}]\ar[dr,bend left,"a^+"]\\
      y\ar[ur,bend left,"a^-"]\ar[rr,bend right,"\id_y"',""{name=I}]&&y
      \ar[from=X,to=I,Rightarrow,"\lambda_a",shorten=3]
    \end{tikzcd}
    \qquad\qquad\qquad\qquad
    \begin{tikzcd}
      &y\ar[phantom,""'{name=X,below=5}]\ar[dr,bend left,"a^-"]\\
      x\ar[ur,bend left,"a^+"]\ar[rr,bend right,"\id_x"',""{name=I}]&&x
      \ar[from=X,to=I,Rightarrow,"\rho_a",shorten=3]
    \end{tikzcd}
  \]
\end{itemize}
As for the definition of $\freegpd P_1$, we would expect some laws to hold (associativity of composition, exchange law, etc.), but those are not needed here because we only consider the mere existence of a 2-cell with given 1-cell as boundary. They would however be required if we were to define 3-polygraphs.

\paragraph{Tietze transformations}
\newcommand{\T}[1]{\textnormal{(T#1)}}
We now define operations on polygraphs, called \emph{Tietze transformations}, turning a 2-polygraph~$P$ into another 2-polygraph~$Q$. Those are of the following three kinds.
\begin{enumerate}[\T0]
\item[\T0] Starting from~$P$ together with $a:P_0$, we construct the polygraph~$Q$ with
  \begin{align*}
    Q_0&=P_0\sqcup\set{\pt}
    &
    Q_1(x,y)&=
    \begin{cases}
      P_1(x,y)&\text{if $x:P_0$ and $y:P_0$}\\
      1&\text{if $x\eqdef a$ and $y\eqdef\pt{}$}\\
      0&\text{otherwise}
    \end{cases}
    \\
    &&
    Q_2(u,v)&=
    \begin{cases}
      P_2(u,v)&\text{if $u,v:\freegpd P_1(x,y)$ for some $(x,y):\Spheres0P$}\\
      0&\text{otherwise}
    \end{cases}
  \end{align*}
\item[\T1] Starting from~$P$ together with $w:\freegpd P_1(a,b)$ for some $(a,b):\Spheres0P$, we construct the polygraph~$Q$ with
  \begin{align*}
    Q_0&=P_0
    &
    Q_1(x,y)&=
    \begin{cases}
      P_1(x,y)\sqcup\set{\pt{}}&\text{if $x\eqdef a$ and $y\eqdef b$}\\
      P_1(x,y)&\text{otherwise}
    \end{cases}
    \\
    &&
    Q_2(u,v)&=
    \begin{cases}
      P_2(u,v)&\text{if $u,v:\freegpd P_1(x,y)$ for some $(x,y):\Spheres0P$}\\
      1&\text{if $u\eqdef w$ and $v\eqdef\pt$}\\
      0&\text{otherwise}
    \end{cases}
  \end{align*}
\item[\T2] Starting from~$P$ together with $\alpha:\freegpd P_2(w,w')$ for some $(w,w'):\Spheres1P$, we construct the polygraph~$Q$ with
  \begin{align*}
    Q_0&=P_0
    &
    Q_1&=P_1
    &
    Q_2(u,v)&=
    \begin{cases}
      P_2(u,v)\sqcup\set{\pt}&\text{if $u\eqdef w$ and $v\eqdef w'$}\\
      P_2(u,v)&\text{otherwise}
    \end{cases}
  \end{align*}
\end{enumerate}
Graphically, we have that
\begin{enumerate}[\T0]
\item[\T0] a transformation \T0 consists, given a 0-generator $a:P_0$, in adding a new 0-generator $\pt$ and a 1-generator from $a$ to~$\pt$:
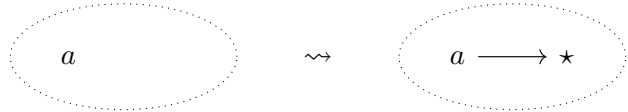
  \[
    \begin{tikzcd}[execute at end picture = {\draw[dotted] (C) ellipse (1.5 and .7);}]
      \ar[r,phantom,""{name=C}]a&{}
    \end{tikzcd}
    \qquad\rightsquigarrow\qquad
    \begin{tikzcd}[execute at end picture = {\draw[dotted] (C) ellipse (1.5 and .7);}]
      \ar[r,phantom,""{name=C}]a\ar[r]&\pt
    \end{tikzcd}
  \]
\item[\T1] a transformation \T1 consists, given a 1-cell $w:\freegpd P_1(x,y)$, in adding a new 1-generator $\pt$ together with a 2-generator from~$w$ to $\pt$:
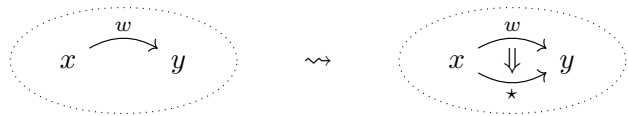
  \[
    \begin{tikzcd}[execute at end picture = {\draw[dotted] (C) ellipse (1.5 and .7);}]
      \ar[r,phantom,""{name=C}]x\ar[r,bend left,"w"]&y
    \end{tikzcd}
    \qquad\rightsquigarrow\qquad
    \begin{tikzcd}[execute at end picture = {\draw[dotted] (C) ellipse (1.5 and .7);}]
      \ar[r,phantom,""{name=C}]x\ar[r,bend left,"w"]\ar[r,bend right,"\pt"']\ar[r,phantom,"\Downarrow"]&y
    \end{tikzcd}
  \]
\item[\T2] a transformation \T2 consists, given a 2-cell $\alpha:\freegpd P_2(w,w')$, in adding a new 2-generator from $w$ to $w'$:
  \[
    \begin{tikzcd}[execute at end picture = {\draw[dotted] (C) ellipse (1.5 and .7);}]
      \ar[r,phantom,""{name=C}]x\ar[r,bend left,"w"]\ar[r,bend right,"w'"']\ar[r,phantom,"{\scriptstyle\alpha\!\!}\Downarrow\phantom{\Downarrow{\scriptstyle\!\!\pt}}"]&y
    \end{tikzcd}
    \qquad\rightsquigarrow\qquad
    \begin{tikzcd}[execute at end picture = {\draw[dotted] (C) ellipse (1.5 and .7);}]
      \ar[r,phantom,""{name=C}]x\ar[r,bend left,"w"]\ar[r,bend right,"w'"']\ar[r,phantom,"{\scriptstyle\alpha\!\!}\Downarrow\Downarrow{\scriptstyle\!\!\pt}"]&y
    \end{tikzcd}
  \]
\end{enumerate}

\paragraph{Tietze equivalence}
Two 2-polygraphs are \emph{Tietze equivalent} when they are related by the smallest equivalence relation identifying two 2-polygraphs related by a Tietze transformation.

\begin{theorem}
  \label{tietze-invariant}
  Two Tietze equivalent 2-polygraphs~$P$ and~$Q$ present the same types.
\end{theorem}
\begin{proof}
  It is enough to show that for each Tietze transformation from~$P$ to~$Q$, the polygraphs present the same type, \ie we have~$\ptype P=\ptype Q$. For transformations \T0 and \T1, we will actually be able to show the stronger property that we have $\gtype P=\gtype Q$.
  \begin{enumerate}[\T0]
  \item[\T0] By \cref{polygraph-colimit}, we have that $\gtype Q$ is the colimit of
    \begin{equation}
      \label{polygraph-colimit-T0}
      \begin{tikzcd}
        &&\G\times\Sigma\Spheres1P.P_2\ar[d]\ar[r,""]\ar[dr,phantom,pos=1,"\ulcorner"]&\Sigma\Spheres1P.P_2\ar[d]\\
        \Sigma\Spheres0P.P_1\sqcup 1\ar[r,shift left,""]\ar[r,shift right,,""']&P_0\sqcup 1\ar[r]&\gtype{Q'}\ar[r]&\gtype{Q}
      \end{tikzcd}
    \end{equation}
    %
    By commutation of colimits (see \cite[Lemma 1.8.3]{brunerie2016homotopy}), we have that $\gtype{Q'}$ is the coequalizer
    \[
      \begin{tikzcd}
        1\ar[r,shift left,"a"]\ar[r,shift right,,"\pt"']&\gtype{P'}\sqcup 1\ar[r,dotted]&\gtype{Q'}
      \end{tikzcd}
    \]
    or equivalently the pushout
    \[
      \begin{tikzcd}
        1\ar[d,"\id_1"']\ar[r,"a"]\ar[dr,phantom,pos=1,"\ulcorner"]&\gtype{P'}\ar[d]\\
        1\ar[r]&\gtype{Q'}
      \end{tikzcd}
    \]
    Moreover, the map $\G\times\Sigma\Spheres1P.P_2\to\gtype{Q'}$ factors through $\gtype{P'}$ (because the boundary of a 2-generator in $P_2$ lies in $\freegpd P_1$, as opposed to $\freegpd Q_1$), so that the definition \cref{polygraph-colimit-T0} reduces to the following sequence of pushouts:
    \[
      \begin{tikzcd}
        &\G\times\Sigma\Spheres1P.P_2\ar[d]\ar[r,""]\ar[dr,phantom,pos=1,"\ulcorner"]&\Sigma\Spheres1P.P_2\ar[d]\\
        1\ar[d,"\id_1"']\ar[r,"a"]\ar[dr,phantom,pos=1,"\ulcorner"]&\gtype{P'}\ar[d]\ar[r]\ar[dr,phantom,pos=1,"\ulcorner"]&\gtype{P}\ar[d]\\
        1\ar[r]&\gtype{Q'}\ar[r]&\gtype{Q}
      \end{tikzcd}
    \]
    We thus have the pushout
    \[
      \begin{tikzcd}
        1\ar[d,"\id_1"']\ar[r,"a"]\ar[dr,phantom,pos=1,"\ulcorner"]&\gtype{P}\ar[d]\\
        1\ar[r]&\gtype{Q}
      \end{tikzcd}
    \]
    from which we deduce that $\gtype P=\gtype Q$ and thus $\ptype P=\ptype Q$.
  \item[\T1] By \cref{polygraph-colimit}, $\gtype Q$ is the colimit
    \[
      \begin{tikzcd}
        &&\G\times(\Sigma\Spheres1P.P_2\sqcup 1)\ar[d]\ar[r,""]\ar[dr,phantom,pos=1,"\ulcorner"]&\Sigma\Spheres1P.P_2\ar[d]\\
        \Sigma\Spheres0P.P_1\sqcup 1\ar[r,shift left,""]\ar[r,shift right,,""']&P_0\ar[r]&\gtype{Q'}\ar[r]&\gtype{Q}
      \end{tikzcd}
    \]
    By a similar reasoning as above on colimits, we obtain that~$\gtype Q$
    is the pushout
    \begin{equation}
      \label{T1-pushout}
      \begin{tikzcd}
        \G\ar[d]\ar[r,"\pt+w"]\ar[dr,phantom,pos=1,"\ulcorner"]&\gtype{P}/(a=b)\ar[d]\\
        1\ar[r]&\gtype Q
      \end{tikzcd}      
    \end{equation}
    where $\gtype P/(a=b)$
    is the pushout
    \[
      \begin{tikzcd}
        2\ar[d]\ar[r,"a+b"]\ar[dr,phantom,pos=1,"\ulcorner"]&\gtype{P}\ar[d]\\
        I\ar[r]&\gtype P/(a=b)
      \end{tikzcd}
    \]
    We write $\pt:a=b$ for the path in $\gtype P/(a=b)$ added by the colimit.
    We can construct the following diagram:
    \[
      \begin{tikzcd}
        2 & I & {\gtype P} \\
        I & \G & {\gtype P/(a=b)}
        \arrow[from=1-1, to=1-2]
        \arrow[from=1-1, to=2-1]
        \arrow["w", from=1-2, to=1-3]
        \arrow[from=1-2, to=2-2]
        \arrow[from=1-3, to=2-3]
        \arrow[from=2-1, to=2-2]
        \arrow["\lrcorner"{anchor=center, pos=0.125, rotate=180}, draw=none, from=2-2, to=1-1]
        \arrow["{\pt + w}"', from=2-2, to=2-3]
      \end{tikzcd}
    \]
    The left square is a pushout by construction, see \cref{globe-pushout}, and the outer rectangle is a pushout because the top map sends the two points of $2$ to $a$ and $b$ respectively, and the bottom one sends the interval to the path $\pt$ in $\gtype P/(a=b)$, see \cref{T1-pushout}.
    Therefore, the right square is also a pushout and, combining this with the definition of $\gtype Q$, we thus have the following pushout diagram:
    \[
      \begin{tikzcd}
        \I\ar[d]\ar[r,"w"]\ar[dr,phantom,pos=1,"\ulcorner"]&\gtype P\ar[d]\\
        \G\ar[d]\ar[r,"\pt+w"]\ar[dr,phantom,pos=1,"\ulcorner"]&\gtype P/(a=b)\ar[d]\\
        1\ar[r]&\gtype Q
      \end{tikzcd}
    \]
    To sum up, replacing~$I$ by~$1$ which is equivalent, $\gtype Q$ can be obtained from $\gtype P$ as the pushout
    \[
      \begin{tikzcd}
        1\ar[d,"\id"']\ar[r,"w"]\ar[dr,phantom,pos=1,"\ulcorner"]&\gtype P\ar[d]\\
        1\ar[r]&\gtype Q
      \end{tikzcd}
    \]
  from which we deduce that $\gtype P=\gtype Q$ and thus $\ptype P=\ptype Q$.

  \item[\T2] By \cref{polygraph-colimit}, $\gtype Q$ is the colimit
    \[
      \begin{tikzcd}
        &&\G\times(\Sigma\Spheres1P.P_2\sqcup 1)\ar[d]\ar[r,""]\ar[dr,phantom,pos=1,"\ulcorner"]&\Sigma\Spheres1P.P_2\ar[d]\\
        \Sigma\Spheres0P.P_1\ar[r,shift left,""]\ar[r,shift right,,""']&P_0\ar[r]&\gtype{Q'}\ar[r]&\gtype{Q}
      \end{tikzcd}
    \]
    By reasoning on colimits, we find that $\gtype Q$ is the pushout
    \begin{equation}
      \label{T2-pushout}
      \begin{tikzcd}
        \G\ar[d]\ar[r,"w+w'"]\ar[dr,phantom,pos=1,"\ulcorner"]&\gtype P\ar[d]\\
        1\ar[r]&\gtype Q
      \end{tikzcd}
    \end{equation}
    and we are in the situation
    \[
      \begin{tikzcd}
        \G\ar[d]\ar[rr,bend left,"w+w'"]\ar[r]\ar[dr,phantom,pos=1,"\ulcorner"]&1\ar[d]\ar[r]&\gtype P\ar[d]\\
        1\ar[r]&\S2\ar[r]&\gtype Q
      \end{tikzcd}
    \]
    where the outer square is a pushout, see \cref{T2-pushout}, and the left square is a pushout (this is the definition of $\S2$ as the suspension of~$\S1$). The square on the right is thus also a pushout.
    Since 1-truncation is left adjoint it preserves pushouts~\cite[Theorem 7.4.12]{hottbook} and we have a pushout diagram
    \[
      \begin{tikzcd}
        1\ar[d]\ar[r]\ar[dr,phantom,pos=1,"\ulcorner"]&\gtrunc{\gtype P}\ar[d]\\
        \gtrunc{\S2}\ar[r]&\gtrunc{\gtype Q}
      \end{tikzcd}
    \]
    which shows that $\gtrunc{\gtype P}=\gtrunc{\gtype Q}$ (since $\gtrunc{\S2}=1$, the left map is an identity).
    \qedhere
  \end{enumerate}
\end{proof}

  % ...

  % ...

\begin{example}
  \label{braid3}
  Consider the fibered 2-polygraph~$P$ with $P_0=\set\pt$, $P_1=\set{a,b}$, $P_2=\set{\beta}$ with
  \begin{align*}
    \partial(\beta)&=(aba,bab)
  \end{align*}
  \ie
  \[
    \begin{tikzcd}[row sep=small]
      &\pt\ar[r,"b",""'{name=b}]&\pt\ar[dr,"a",bend left]\\
      \pt\ar[ur,"a",bend left]\ar[dr,"b"',bend right]&&&\pt\\
      &\pt\ar[r,"a"',""{name=a}]&\pt\ar[ur,"b"',bend right]
      \ar[from=b,to=a,Rightarrow,"\beta"',shorten=7]
    \end{tikzcd}
  \]
  Using a more traditional notation for presentations, this polygraph can be noted as
  \[
    P
    \qdefd
    \Pres{\pt}{a,b}{aba=bab}
  \]
  It is known that $\ptype P$ is a delooping of $B_3$, the group of braids with three strands. We have the following series of Tietze transformations:
  \begin{enumerate}[\T2]
  \item[\T1] we add a 1-generator $c:\pt=\pt$ and a 2-generator $c=ba$:
    \[
      \Pres{\pt}{a,b,c}{aba=bab,c=ba}
    \]
  \item[\T2] we add the derivable relation $ac=cb$:
    \[
      \Pres{\pt}{a,b,c}{aba=bab,ac=cb,c=ba}
    \]
  \item[\T2] (backward), we remove the derivable relation $aba=bab$:
    \[
      \Pres{\pt}{a,b,c}{ac=cb,c=ba}
    \]
  \end{enumerate}
  We finally deduce by \cref{tietze-invariant} that this last 2-polygraph also presents a delooping of $B_3$.
\end{example}

\begin{remark}
  \label{knuth-bendix}
  The Knuth-Bendix completion algorithm~\cite{knuth1983simple}, which iteratively adds relations in order to hopefully compute a presentation that is convergent (both confluent and terminating), is also based on Tietze transformations $\T2$, see~\cite{polygraphs}. This explains why it also preserves the presented type. For instance, it can be used to compute the following convergent presentation of the group~$B_3$ of \cref{braid3}:
  \[
    \Pres{\pt}{a,b,c}{ccb=acc, bcb=cc, ca=bc, baca=cac, bab=aba, ab=c}
  \]
  However, note that we first need to use a transformation $\T1$ in order to add the 1-generator~$c$, without it no convergent presentation of $B_3$ can be found (on the generators $a$ and $b$ only)~\cite{kapur1985finite}.
\end{remark}

\begin{remark}
  \label{twoard-completeness}
  In group theory~\cite{tietze1908topologischen} (as well as in polygraph theory~\cite{polygraphs}), it is known that Tietze transformations are \emph{complete} in the sense that we have a converse to the correctness property (\ie the analogue of \cref{tietze-invariant}): if two presentations present the same group then they are Tietze equivalent.
  %
  We expect that a similar property can be shown for 2-polygraphs, but this is left for future works. A proof of completeness in the case of 1-polygraphs can be found in~\cite{coh-tietze}.
\end{remark}



\section{Cayley graphs}
\label{cayley}
Throughout the section, we fix a group~$G$ together with a generating set~$X$. We have seen in \cref{delooping-hit} that a delooping of~$G$ can be obtained by further homotopy quotienting a delooping of~$\freegroup X$. The kernel of the map $\freegroup\gamma:\freegroup X\to G$ measures the defect of~$\freegroup X$ from being~$G$, which corresponds to the relations of the group. We show here that, under the delooping operation, those relations are precisely encoded by the Cayley graph~\cite{cayley1878desiderata,lyndon1977combinatorial}, a classical and useful construction in group theory which can be associated to any generated group.

The \emph{Cayley graph} of~$G$, with respect to the generating set~$X$, is the directed graph whose vertices are the elements of~$G$, and such that for every vertex $g:G$ and generator $x:X$, we have an edge $g\to gx$. In homotopy type theory, it is thus natural to represent it as the higher inductive type $\Cayley XG$ defined as
\[
  \begin{array}{l@{\ :\ }l}
    \cvertex&G\to\Cayley XG\\
    \cedge&(g:G)(x:X)\to\cvertex g=\cvertex(gx)
  \end{array}
\]

\begin{example}
  Consider the cyclic group $\Z_5$ (with $2$ as generator) and the dihedral group $D_5$ (with $r$ and $s$ as generators, see also \cref{Dn}) which can be presented by
  \begin{align*}
    \Z_5&=\pres{a}{a^5=1}
    &
    D_5&=\pres{r,s}{r^5=1,s^2=1,rsrs=1}
  \end{align*}
  The associated Cayley graphs are respectively
  
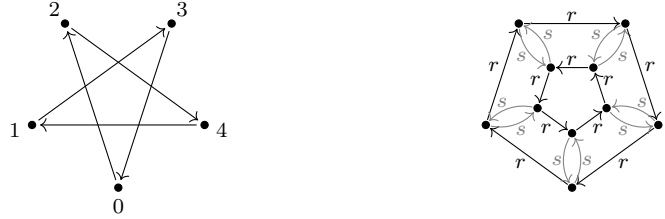
\begin{align*}
    \begin{tikzpicture}[baseline=(b.base),scale=1.2,every edge/.append style = {shorten <=3,shorten >=3}]
      \coordinate (b) at (0,0);
      \foreach \i in {0,...,4} {
        \filldraw ({(\i+3/4)*360/5}:1) circle (.04);
        \draw ({((-\i+3)+3/4)*360/5}:1.2) node {$\scriptstyle\i$};
        \draw[shorten <=2,shorten >=2] ({(\i+3/4)*360/5}:1) edge[<-] ({(\i+2+3/4)*360/5}:1);
      }
    \end{tikzpicture}
    &&
    \begin{tikzpicture}[baseline=(b.base),scale=1.2,every edge/.append style = {shorten <=2,shorten >=2}]
      \coordinate (b) at (0,0);
      \foreach \i in {0,...,4} {
        \filldraw ({(\i+3/4)*360/5}:1) circle (.04);
        \filldraw ({(\i+3/4)*360/5}:.4) circle (.04);
        \draw ({(\i+3/4)*360/5}:1) edge[<-,inner sep=0.2ex,"$\scriptstyle r$"'] ({(\i+1+3/4)*360/5}:1);
        \draw ({(\i+3/4)*360/5}:.4) edge[->,inner sep=0.2ex,"$\scriptstyle r$"'] ({(\i+1+3/4)*360/5}:.4);
        \draw ({(\i+3/4)*360/5}:1) edge[->,bend right,gray,inner sep=0.1ex,"$\scriptstyle s$"'] ({(\i+3/4)*360/5}:.4);
        \draw ({(\i+3/4)*360/5}:.4) edge[->,bend right,gray,inner sep=0.1ex,"$\scriptstyle s$"'] ({(\i+3/4)*360/5}:1);
      }
    \end{tikzpicture}    
  \end{align*}
\end{example}

All the previous constructions can be reformulated in terms of polygraphs. First note that the generating set can be understood as a 1-polygraph~$P$ with $P_0\defd 1$ and $P_1\defd X$, so that $\B\freegroup X$ is precisely~$\gtype P$. Similarly, from their very definition, Cayley graphs are generated by polygraphs. Namely, it can be observed that

\begin{proposition}
  \label{cayley-1pol}
  Consider the fibered 1-polygraph~$Q$ with
  \begin{align*}
    Q_0&\defd G
    &
    Q_1&\defd G\times X
    &
    \partial:Q_1&\to \Spheres0Q
    \\
    &&&&(g,x)&\mapsto (g,gx)
  \end{align*}
  The Cayley graph of~$G$ is precisely its generated type:
  \[
    C(X,G)
    \qeq
    \gtype Q
  \]
\end{proposition}

\noindent
Our main result in this section is that Cayley graphs satisfy the following property.

\begin{theorem}[\agda{Cayley}{Cayley-ker}]
  \label{flattening-cayley}
  The type $\Cayley XG$ is the kernel of the function
  \[
    \B\freegroup \gamma:\B\freegroup X\to\B G
  \]
  induced by~$\gamma$, \ie we have
  \[
    \Cayley XG
    \qeq
    \Sigma(x:\B\freegroup X).(\pt=\B\freegroup\gamma(x))
    \text.
  \]
\end{theorem}
\begin{proof}
  We consider the type family
  \begin{align*}
    F:\B\freegroup X&\to\U\\
    x&\mapsto(\pt=\B\freegroup \gamma(x))
  \end{align*}
  Remember that $\B\freegroup X$ admits a description as a coequalizer, see \cref{wedge-of-circles} and \cref{bouquet-loop-space}:
  \[
    \begin{tikzcd}
      X\ar[r,shift left]\ar[r,shift right]&1\ar[r,dotted]&\B\freegroup X
    \end{tikzcd}
  \]
  Hence, by the flattening lemma for coequalizers (see \cref{flattening-coequalizers} and \cite[Section~6.12]{hottbook}), we have a coequalizer of total spaces
  \[
    \begin{tikzcd}
      \Sigma X.F(\pt)
      \ar[r,shift left,"{(x,p)\mapsto(\pt,p)}"]
      \ar[r,shift right,"{(x,p)\mapsto(\pt,\subst Fp(x))}"']
      &[50]
      \Sigma 1.F(\pt)
      \ar[r,dotted]
      &
      \Sigma \B\freegroup X.F
    \end{tikzcd}
  \]
  By using the properties of transport in path
  spaces~\cite[Theorem~2.11.3]{hottbook}, it can be shown that the bottom map
  sends $(x,p)$ to $(\pt,p\pcomp\B\freegroup\gamma(x))$. Moreover,
  $\B\freegroup \gamma$ is pointed, so $F(\pt)$ is equal to $\Loop\B G$,
  \ie $G$, and we have the following coequalizer:
  \[
    \begin{tikzcd}
      X\times G
      \ar[r,shift left,"{(x,g)\mapsto g}"]
      \ar[r,shift right,"{(x,g)\mapsto gx}"']
      &[20]
      G
      \ar[r,dotted]
      &
      \Sigma \B\freegroup X.F
    \end{tikzcd}
  \]
  It follows that $\Sigma\B\freegroup X.F$ consists of $\sizeof G$ points and a
  path $g=gx$ for each pair $(x,g):X\times G$, and is therefore equal to the
  Cayley graph $\Cayley XG$.
\end{proof}

\noindent
The above result can be interpreted as stating that we have a fiber sequence
\[
  \begin{tikzcd}
    \Cayley XG\ar[r] & \B\freegroup X \ar[r,"\B\freegroup\gamma"]&\B G
  \end{tikzcd}
\]
Under the Grothendieck duality, the map $\B\freegroup\gamma:\B\freegroup X\to\B G$ corresponds to a type family $\phi:\B G\to\U$ with $\phi\,(\pt)=\Cayley XG$, which means that it encodes an action of~$G$ on the Cayley graph (which is the canonical one, sending $h\cdot g$ to $hg$). Moreover, we have $\B\freegroup X=\Sigma(\B G).\phi$, meaning that $\B\freegroup X$ can be understood as the homotopy quotient of the Cayley graph under this action. This point of view is developed in~\cite[Proposition~16]{lens}.
Moreover, the map $\B\freegroup X$ is an approximation of $\B G$ in the following sense:

\begin{proposition}[\agda{Cayley}{Cayley-connected}]
  \label{cayley-connected}
  The Cayley graph $\Cayley XG$ is connected and the map $\B\freegroup X\to\B G$ is thus $0$-connected.
\end{proposition}
\begin{proof}
  We consider $\Cayley XG$ as pointed by the point corresponding to the neutral element of~$G$.
  It is enough to show that for every point $x:\Cayley XG$ there merely exists a path $\pt=x$, which is easily done by induction on~$x$ using the fact that $X$ is generating.
\end{proof}

\paragraph{Relations}
The long exact sequence of homotopy groups induced by the above fiber sequence~\cite[Theorem~8.4.6]{hottbook} implies in particular that we have the following short exact sequence of groups
\[
  \begin{tikzcd}
  \label{exseq-cayley}
    1\ar[r] &\Loop\Cayley XG\ar[r] &\freegroup X \ar[r]&G\ar[r] & 1
  \end{tikzcd}
\]
which shows that $\Loop\Cayley XG$ is the (free) group encoding relations of~$G$ with respect to~$X$. Indeed, we have $\Cayley XG=\B\freegroup R$, where $R$ is a choice of $\sizeof G\times(\sizeof X-1)+1$ relations: these are the loops in the Cayley graph after contracting $\sizeof G-1$ edges to obtain a wedge of circles. In some sense, \cref{flattening-cayley} provides an internalization of the fact that~$G$ is presented by $\pres XR$, contrasting with the point of view developed in \cref{delooping-hit}.

\paragraph{The Cayley complex}
\newcommand{\CC}{\operatorname{C}}
We now explain that we can extend the previous construction in higher dimensions in order to define internally a type corresponding to the classical \emph{Cayley complex}~\cite{lyndon1977combinatorial}.

Consider a group~$G$ together with a presentation~$\pres XR$, and write $P$ for the 2-polygraph corresponding to this presentation (by \cref{pres-pol}). We write $\B_2P$ for the $2$-skeleton of the type $\B P$ defined in \cref{delooping-hit}, \ie the higher inductive type with generators
\[
  \begin{array}{l@{\ :\ }l}
    \pt&{\B P}\\
    \cgen&X\to(\pt=\pt)\\
    \crel&(r:R)\to(\freegroup\cgen(\fst(r))=\freegroup\cgen(\snd(r)))\\
  \end{array}
\]
Alternatively, $\B_2P$ is precisely the type $\gtype P$ generated by the 2-polygraph~$P$. This type can be considered as an approximation of~$\B P$: it is ``almost'' $\B P$, except that it lacks the groupoid truncation. In particular, this type has the same first two homotopy groups as $\B P$. Below, we write $\B_1P$ for the type generated by the 1-polygraph of~$P$ (note that if we write $X\defd P_1$, then $\B_1P$ is simply another notation for the type $\B\freegroup X$ studied in the previous section).

\begin{lemma}
  \label{B2P-pi01}
  The type $\B_2 P$ is connected and we have $\pi_1(\B_2P)=G$.
\end{lemma}
\begin{proof}
  By definition, the type $\B_2P$ is a quotient of the type $\B_1P$ (under the relations specified by $\crel$, corresponding to $P_2$). Since the latter is connected by \cref{cayley-connected}, the former is also connected.
  Moreover, since $\B G$ is the groupoid truncation of $\B_2P$, they have the same fundamental group. Namely,
  \[
    \pi_1(\B_2P)
    \qeqdef\quad
    \strunc{\pt\overset{\B_2P}=\pt}
    \qeq
    (\gtrunq\pt\overset{\gtrunc{\B_2P}}=\gtrunq\pt)
    \qeqdef
    \pi_1(\B G)
    \qeq
    G
  \]
  where the first non-definitional equality is due to~\cite[Theorem 7.3.12]{hottbook}.
\end{proof}

We would like to ``measure the difference'' between the two types $\B_2 P$ and $\B G$. Consider the canonical map
\[
  \phi:\B_2P\to\B G
\]
which is given by sending every path $\cgen x$ to the loop in $\B G$ corresponding to~$x$ and each cell $\crel r$ to the cell in $\B G$ witnessing that the relation $r$ is satisfied in~$G$. Alternatively, this map is also induced by the fact that we have a definition of $\B G$ as a higher inductive type with more constructors than $\B_2P$ (namely the constructor $\ctrunc$, which corresponds to taking the groupoid truncation), \ie we have $\B G=\gtrunc{\B_2 P}$ and $\phi$ is the map
\begin{equation}
  \label{phi-gtrunc}
  \gtrunq{{-}}:\B_2P\to\gtrunc{\B_2G} 
\end{equation}

\begin{definition}
  The \emph{Cayley complex} $\CC P$ associated to the presentation~$P$ is the inductive type defined by
  \[
    \begin{array}{l@{\ :\ }l}
      \cvertex&G\to\CC P\\
      \cedge&(g:G)(x:X)\to\cvertex g=\cvertex(gx)\\
      \ccell&(g:G)(r:R)\to\freegroup{(\cedge g)}(\fst(r))=\freegroup{(\cedge g)}(\snd(r))
    \end{array}
  \]
  where, given $g:G$ and $u\defd x_1x_2\ldots x_n:\freegroup X$, we have that $\freegroup{(\cedge g)}\,u$ is the path
  \[
    \begin{tikzcd}[column sep=small]
      \cv g
      \ar[r,equals,"\ce g\,x_1"]
      &[5]
      \cv(gx_1)
      \ar[r,equals,"\ce(gx_1)\,x_2"]
      &[20]
      \cv(gx_1x_2)
      \ar[r,equals,"\ldots"]
      &
      \cdots
      \ar[r,equals,"\ldots"]
      &
      \cv(gx_1\ldots x_{n-1})
      \ar[r,equals,"\ce g(x_1\ldots x_{n-1})\,x_n"]
      &[45]
      \cv(gx_1\ldots x_n)
    \end{tikzcd}
  \]
  where $\cv$ (\resp $\ce$) is a short notation for $\cvertex$ (\resp $\cedge$).  
\end{definition}

\begin{proposition}
  The Cayley complex $\CC P$ is the type $\gtype Q$ generated by the fibered 2-polygraph~$Q$ with
  \begin{align*}
    Q_0&\defd G
    &
    Q_1&\defd G\times X
    &
    Q_2&\defd G\times R
  \end{align*}
  and the expected boundary maps.
\end{proposition}


\noindent
The fiber sequence of \cref{flattening-cayley} generalizes to this setting:

\begin{theorem}
  \label{cayley-complex-fiber-sequence}
  We have a fiber sequence
  \[
  \begin{tikzcd}
    \CC P\ar[r]&\B_2 P\ar[r,"\phi"]&\B G
  \end{tikzcd}
  \]
\end{theorem}
\begin{proof}
  The main arguments of the proof are the following ones, essentially following the proof of \cref{flattening-cayley}. A more detailed version of this proof (although in the case where $G$ is the quaternion group~$Q$, but not depending on thisin an essential way) can be found in~\cite[Section 3.5]{hypercubical}.

  We write $P'$ for the underlying 1-polygraph of the 2-polygraph~$P$ and $\B_1P\defd\gtype{P'}$ for the generated type.
  We have seen in \cref{2pol-gtype-pushout} that the type $\B_2P\eqdef\gtype P$ can be obtained from the type $\B_1P$ as the pushout
  \[
    \begin{tikzcd}
      \G\times\Sigma\Spheres1P.P_2\ar[d,"f"']\ar[r,"p"]\ar[dr,phantom,pos=1,"\ulcorner"]&\Sigma\Spheres1P.P_2\ar[d,"g"]\\
      \B_1P\ar[r,"q"']&\B_2P
    \end{tikzcd}  
  \]
  where the map $f$ essentially sends a 2-generator to the sphere in $\B_1P$ corresponding to its boundary and $p$ is the second projection.
  Now, consider the type family
  \begin{align*}
    F:\B_2P&\to\U\\
    x&\mapsto(\pt=\phi(x))
  \end{align*}
  By the flattening lemma for pushouts (\cref{flattening-pushouts}), we thus have a pushout of total spaces of the type family~$F$:
  \[
    \begin{tikzcd}[column sep=huge]
      \Sigma(\G\times\Sigma\Spheres1P.P_2).F\circ g\circ p\ar[d,"\Sigma f.e"']\ar[r,"\Sigma p.(\lambda\_.\id)"]\ar[dr,phantom,pos=1,"\ulcorner"]&\Sigma(\Sigma\Spheres1P.P_2).F\circ g\ar[d,"\Sigma g.(\lambda\_.\id)"]\\
      \Sigma(\B_1P).F\circ q\ar[r,"\Sigma q.(\lambda\_.\id)"']&\Sigma(\B_2P).F
    \end{tikzcd}
  \]
  where $e:(x:\G\times\Sigma\Spheres1P.P_2)\to F\circ g\circ p(x)\to F\circ q\circ f(x)$ is the map canonically induced by the identity $g\circ p=q\circ f$.
  By definition, the type $\Sigma(\B_2P).F$ on the lower right is the kernel of $\phi$.  
  Remembering that $\B_1P$ is another notation for $\B\freegroup X$, we have by \cref{flattening-cayley} that the type $\Sigma(\B_1P).F\circ q$ at the lower left is the Cayley graph $\Cayley XG$. Namely, an element of $\Sigma(\B_1P).F\circ q$ consists of an element of $\B_1P$ (\ie a point in the loop corresponding to an element of $\freegroup X$) and an element of~$G$ (because $F\circ q(\pt)$ is $\pt=\pt$ in $\B_2P$, \ie $G$).
  Now, an element of $\Sigma(\G\times\Sigma\Spheres1P.P_2).F\circ g\circ p$ consists in a point $s:\G$, a 2-generator of~$P$ in some 1-sphere $(u,v)$ together with an element~$a$ of~$G$, and the map $\Sigma f.e$ respectively sends the northern and southern hemisphere of $\G$ to the paths corresponding to $u$ and $v$ starting from~$a$ in $\Cayley XG$, and the pushout states that these should be contracted to a point (which is equivalent to filling in the corresponding sphere in $\Cayley XG$ with a disk \cite[Section 6.7]{hottbook}).
\end{proof}

\begin{example}
  \label{quaternion-cayley}
  Consider the quaternion group, which can be presented as
  \[
    Q
    =
    \pres{i,j}{i=jij,j=iji}
  \]
  It is shown in~\cite{hypercubical} that its Cayley graph is
  
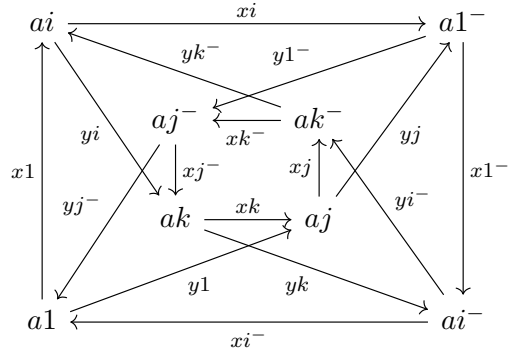
\[
    \begin{tikzcd}
      ai\ar[rrr,"xi"]\ar[ddr,"yi"']&&&\ar[dll,"y1^-"']a1^-\ar[ddd,"x1^-"]\\
      &aj^-\ar[ddl,"yj^-"']\ar[d,"xj^-"]&\ar[l,"xk^-"]\ar[ull,"yk^-"']ak^-\\
      &ak\ar[drr,"yk"']\ar[r,"xk"]&\ar[u,"xj"]aj\ar[uur,"yj"']\\
      a1\ar[urr,"y1"']\ar[uuu,"x1"]&&&\ar[lll,"xi^-"]\ar[uul,"yi^-"']ai^-
    \end{tikzcd}
  \]
  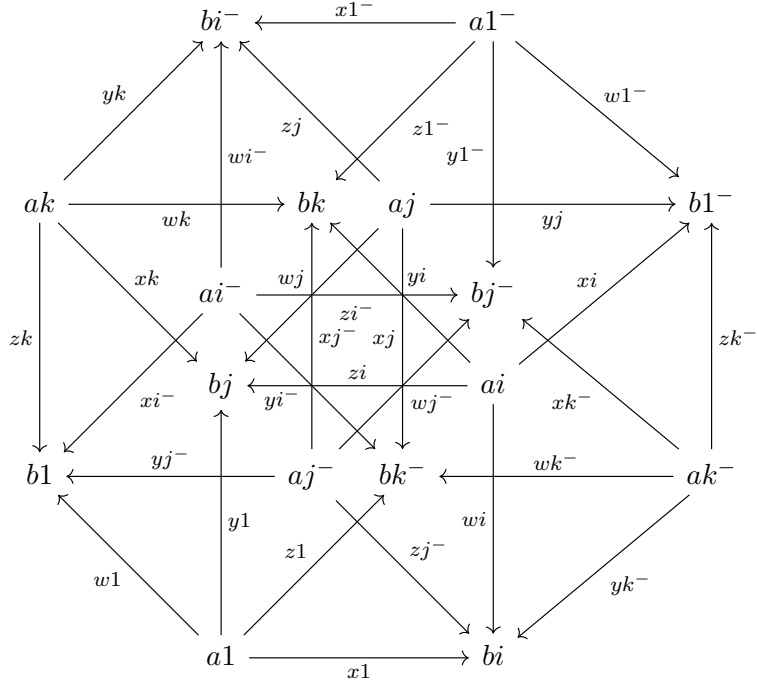
\begin{figure}[t]
    \centering
    \[
  \begin{tikzcd}[column sep={12mm,between origins},row sep={12mm,between origins}]
    &&bi^-&&&a1^-\ar[lll,"x1^-"']\ar[ddll,"z1^-"]\ar[ddd,"y1^-"']\ar[ddrr,"w1^-"]\\
    \\
    ak\ar[ddd,"zk"']\ar[uurr,"yk"]\ar[rrr,"wk"']\ar[ddrr,"xk"]&&&bk&aj\ar[uull,"zj"]\ar[ddll,"wj"']\ar[ddd,"xj"']\ar[rrr,"yj"']&&&b1^-\\
    &&ai^-\ar[ddll,"xi^-"]\ar[uuu,"wi^-"']\ar[rrr,"zi^-"']\ar[ddrr,"yi^-"']&&&bj^-\\
    &&bj&&&ai\ar[uull,"yi"']\ar[lll,"zi"']\ar[ddd,"wi"']\ar[uurr,"xi"]\\
    b1&&&aj^-\ar[lll,"yj^-"']\ar[uuu,"xj^-"']\ar[uurr,"wj^-"']\ar[ddrr,"zj^-"]&bk^-&&&ak^-\ar[uull,"xk^-"]\ar[lll,"wk^-"']\ar[ddll,"yk^-"]\ar[uuu,"zk^-"']\\
    \\
    &&\ar[uull,"w1"]a1\ar[uurr,"z1"]\ar[uuu,"y1"']\ar[rrr,"x1"']&&&bi
  \end{tikzcd}
\]
    \caption{Cayley complex of the quaternion group.}
    \label{tesseract}
  \end{figure}
  and its Cayley complex is shown in \cref{tesseract} (here, all the squares are filled).
\end{example}

\begin{proposition}
  \label{cayley-complex-connected}
  The Cayley complex is 1-connected and the map $\phi:\B_2P\to\B G$ is thus 1\nbd-connec\-ted.
\end{proposition}
\begin{proof}
  Writing~$\pt$ for the point corresponding to the neutral element of the group, we have to show that the type~$\strunc{\pt=x}$ is contractible, for any vertex $x:G$. Since the Cayley complex is connected as a quotient of a connected space (\cref{cayley-connected}) and being contractible is a proposition, we can suppose that we have a path $\pt=x$ and therefore, by path induction, we are left with showing that $\strunc{\pt=\pt}$, \ie $\pi_1(\CC P)$ is contractible.
  By~\cite[Theorem~8.4.6]{hottbook}, the fiber sequence of \cref{cayley-complex-fiber-sequence} induces an exact sequence of homotopy groups
  \[
    \begin{tikzcd}[column sep={25mm,between origins}]
      \pi_2(\B G)\ar[r]&\pi_1(\CC P)\ar[r]&\pi_1(\B_2P)\ar[r]&\pi_1(\B G)
    \end{tikzcd}
  \]
  We have $\pi_2(\B G)=0$ because $\B G$ is a groupoid, $\pi_1(\B_2P)=G$ by \cref{B2P-pi01} and $\pi_1(\B G)=G$ by definition of deloopings, so that the exact sequence can be simplified to
  \[
    \begin{tikzcd}[column sep={25mm,between origins}]
      1\ar[r]&\pi_1(\CC P)\ar[r]&G\ar[r]&G
    \end{tikzcd}
  \]
  The arrow on the right, which is $\pi_1(\phi)$, is easily seen to be the identity, and the exact sequence states that $\pi_1(\CC P)$ is the kernel of this map, \ie $1$, and is thus contractible.
\end{proof}

The previous proposition allows us to show that $\CC P$ can be characterized as the universal covering of $\B_2P$. We recall from~\cite{harper2018covering,galois} that a pointed map $f:A\pto B$ is a \emph{covering} when its fibers are sets. It is a \emph{universal covering} when, moreover, $A$ is 1-connected.

\begin{proposition}
  The map $\CC P\to\B_2P$ exhibits $\CC P$ as the universal covering of $\B_2P$.
\end{proposition}
\begin{proof}
  We have seen in \cref{cayley-complex-connected} that the type $\CC P$ is connected; it remains to be shown that its fibers are sets. By \cite[Section~8.4]{hottbook}, the fiber sequence of \cref{cayley-complex-fiber-sequence} extends on the left as a fiber sequence
  \[
    \begin{tikzcd}
      \Loop\B G\ar[r]&\CC P\ar[r]&\B_2P
    \end{tikzcd}
  \]
  Since we have $\Loop\B G=G$ by definition of~$\B G$, we deduce that the kernel of the map $\CC P\to\B_2P$ is $G$ and thus a set. Since $\B_2P$ is connected and being a set is a proposition~\cite[Theorem 7.1.10]{hottbook}, we deduce that all its fibers are sets.
\end{proof}

\noindent
In fact, the above situation is very generic. We have observed in \cref{phi-gtrunc} that the map $\phi:\B_2P\to\B G$ is the groupoid truncation. More generally, given a pointed connected type~$A$, the map $\gtrunc{{-}}:A\to\gtrunc{A}$ corresponds to the \emph{Galois fibration} associated to~$A$, and its kernel is the universal covering~$\tilde A$ of~$A$, \ie we have a fiber sequence
\[
  \begin{tikzcd}
    \tilde A\ar[r]&A\ar[r,"\gtrunq{{-}}"]&\gtrunc{A}
  \end{tikzcd}
\]
which encodes the canonical action of the fundamental group of~$A$ on its universal covering. This will be detailed in~\cite{galois}.

\paragraph{Higher Cayley complexes}
We expect that the previous constructions extend to higher dimensions. Namely, we should be able to define a notion of \emph{$n$-polygraph} for every natural number~$n$: essentially, an $(n{+}1)$-polygraph consists of an $n$-polygraph~$P$ together with a function $\Spheres nP\to\Set$ which associates to every $n$-sphere in~$P$ a set of \emph{$(n{+}1)$-generators}. The difficulty here lies in the fact that we need to define the notion of $n$-sphere, which in turn requires defining a notion of \emph{$n$-cell} in~$P$ in a coherent way. We leave this for future work. With an $(n{+}1)$-polygraph~$P$, we expect to have a notion of generated type $\gtype P$, and we say that $P$ is a \emph{presentation} of a group~$G$ when $\trunc{n}{\gtype P}=\B G$. Finally, we could define the Cayley complex $\CC P$ associated to~$P$ as the kernel of the truncation map $\trunq{n}{{-}}:\gtype P\to\trunc{n}{\gtype P}$ so that we have a fiber sequence
\[
  \begin{tikzcd}
    \CC P\ar[r]&\gtype P\ar[r]&\B G
  \end{tikzcd}
\]
generalizing \cref{flattening-cayley,cayley-complex-fiber-sequence}. Note that this would exhibit $\CC P$ as the universal $(n{-}1)$\nbd-covering of $\gtype P$, in the sense developed in~\cite{galois}. In particular, \cref{flattening-cayley,cayley-complex-fiber-sequence} respectively exhibit the Cayley graph and the Cayley complex as the universal covering and universal 1-covering of the type~$\gtype P$ generated by the presentation~$P$ of a given group.

This resolution-like process can be iterated in order to obtain better and better approximations $\B_nP$ of $\B G$, and higher Cayley complexes as the fibers of the canonical maps $\B_nP\to\B G$. Moreover, the join construction~\cite{rijke2017join,rijke2018classifying} provides a way to automate this task, see for instance~\cite{buchholtz2017real,lens}.

\section{Future works}
\label{conclusion}
We have presented two ways to improve the known constructions of deloopings of groups when a presentation of the group is given.
%
This work is part of a larger investigation of models of group deloopings that are ``efficient'' in the sense that they allow the use of traditional techniques from group presentations, such as Tietze transformations or coherence based on rewriting~\cite{polygraphs,kraus2022rewriting}. A few instances have already been studied in detail, but many group deloopings remain to be explored. For instance, a construction of infinite real projective space was introduced by Buchholtz and Rijke~\cite{buchholtz2017real}, providing a cellular description of $\B\Z_2$; we have recently been defining \emph{lens spaces}~\cite{lens}, which provide small cellular deloopings of $\B\Z_n$; and we are currently investigating hypercubical manifolds~\cite{hypercubical}, which provide cellular deloopings for the quaternion group (see also \cref{quaternion-cayley}).
More generally, the formalization of group theory in univalent foundations is still under active investigation~\cite{symmetry}, and we aim to develop general techniques for constructing efficient representations of (internal) groups in homotopy type theory, which would open the way to cohomological computations for groups~\cite{brunerie2022synthetic,buchholtz2020cellular,cavallo2015synthetic,lamiaux2023computing,ljungstrom2024computational} and to the definition of group actions on higher types (as a generalization of group actions on sets).
Finally, we have seen that higher inductive types play, in homotopy type theory, a role analogous to that of polygraphs for strict higher categories. Numerous techniques have been developed for polygraphs~\cite{polygraphs}, notably based on rewriting, and we would like to adapt them in this setting. The notion of 2-polygraph has been defined in homotopy type theory by Kraus and von Raumer~\cite{kraus2022rewriting}, who also explained how to obtain coherence (formally, a homotopy basis) from a convergent presentation. In \cite{coh-tietze}, we develop the notion of Tietze transformation for 1-polygraphs and show both correctness and completeness, and in this paper we have developed a similar notion for 2-polygraphs. A general definition of $n$-polygraphs is still missing and is the subject of ongoing work. Having a notion of 3-polygraph would already be a major improvement, since it would allow the manipulation of coherent presentations of groups.


\bibliographystyle{plainurl}
\bibliography{papers}
\end{document}